\documentclass[12pt]{article}
\usepackage{amssymb}
\usepackage{amsmath}
\numberwithin{equation}{section}

\newcommand{\nonu}{\nonumber \\ }

\newcommand{\hs}[1]{\hspace{#1 mm}}

\newcommand{\lda}{\lambda}

\newcommand{\bbet}{{\overline{\beta}}}

\newcommand{\bPhi}{{\overline{\Phi}}}

\newcommand{\bvarphi}{\overline{\varphi}}
\newcommand{\Phidag}{\Phi^\dagger}
\newcommand{\blambda}{{\overline{\lambda}}}

\newcommand{\adag}{a^\dagger}

\newcommand{\ca}{\mbox{$\cal{A}$}}

\newcommand{\cc}{\mbox{$\cal{C}$}}
\newcommand{\cd}{\mbox{$\cal{D}$}}
\newcommand{\ce}{\mbox{${\cal E}$}}
\newcommand{\cf}{\mbox{${\cal F}$}}
\newcommand{{\cg}}{\mbox{$\cal{G}$}}
\newcommand{\ch}{\mbox{$\cal{H}$}}

\newcommand{\cl}{\mbox{$\cal{L}$}}

\newcommand{\cs}{\mbox{$\cal{S}$}}
\newcommand{\cR}{\mbox{$\cal{R}$}}
\newcommand{\ct}{\mbox{$\cal{T}$}}

\newcommand{\prt}{\partial}

\newcommand{\mb}[1]{\hs{4}\mbox{#1}\hs{4}}

\newcommand{\mf}{\mathfrak{f}}
\newcommand{\mg}{\mathfrak{g}}
\newcommand{\mh}{\mathfrak{h}}
\newcommand{\mS}{\mathfrak{S}}
\newcommand{\cinf}{C_0^{\infty}(\RR)}

\newcommand{\cinfa}{C_0^{\infty}(\RR^\alpha)}
\newtheorem{theo}{Theorem}[section]
\newtheorem{prop}[theo]{Proposition}
\newtheorem{defi}[theo]{Definition}

\newtheorem{lem}[theo]{Lemma}

\newcommand{\prf}{\underline{Proof:}\ }
\newcommand{\finprf}{\null \hfill {\rule{5pt}{5pt}}\\[2.1ex]\indent}
\newcommand{\ie}{{\it i.e.}\ }
\newcommand{\CC}{{\mathbb C}}
\newcommand{\RR}{{\mathbb R}}

\newcommand{\ZZ}{{\mathbb Z}}

\newcommand{\1}{\mbox{\hspace{.0em}1\hspace{-.24em}I}}

\newcommand{\topa}[2]{\genfrac{}{}{0pt}{}{#1}{#2}}
\begin{document}
\renewcommand{\thefootnote}{\fnsymbol{footnote}}

\newpage
\pagestyle{empty} \setcounter{page}{0}

\markright{\today\dotfill DRAFT RT-NLS\dotfill }

\newcommand{\LAP}{LAPTH}
\def\logo{{\bf {\huge LAPTH}}}

\centerline{\logo}

\vspace {.3cm}

\centerline{{\bf{\it\Large Laboratoire d'Annecy-le-Vieux de
Physique Th\'eorique}}}

\centerline{\rule{12cm}{.42mm}}

\vspace{20mm}

\begin{center}

   {\LARGE  {\sffamily Solving the quantum non-linear Schr\"odinger
   equation\\[1.2ex] with $\delta$-type impurity }}\\[1cm]

\vspace{10mm}

{\large V. Caudrelier$^a$\footnote{caudreli@lapp.in2p3.fr},
  M. Mintchev$^b$\footnote{mintchev@mail.df.unipi.it} and E.
Ragoucy$^a$\footnote{ragoucy@lapp.in2p3.fr}}\\[.21cm]
  $^a$ Laboratoire de Physique Th{\'e}orique \LAP\footnote{UMR 5108
     du CNRS, associ{\'e}e {\`a} l'Universit{\'e} de
Savoie.}\\ 
     LAPP, BP 110, F-74941  Annecy-le-Vieux Cedex, France.\\[.242cm]
  $^b$ INFN and Dipartimento di Fisica, Universit\'a di
      Pisa,\\ Via Buonarroti 2, 56127 Pisa, Italy
\end{center}
\vfill

\begin{abstract}
We establish the exact solution of the nonlinear Schr\"odinger
equation with a delta--function impurity, representing a point--like
defect which reflects and transmits. We solve the problem both at
the classical and the second quantized levels.
In the quantum case the Zamolodchikov-Faddeev algebra,
familiar from the case without impurities, is substituted by the
recently discovered reflection--transmission (RT) algebra, which
captures both particle--particle and particle--impurity interactions.
The off--shell quantum solution is expressed in terms of the generators
of the RT algebra and the exact scattering matrix of the theory is derived.
\end{abstract}
\vfill \centerline{PACS numbers: 02.30.Ik}
\vfill
\centerline{J. Math. Phys. to appear}
\vfill
\rightline{\tt math-ph/0404047}
\rightline{\LAP-1038/04}
\rightline{IFUP-TH 16/2004}
\rightline{April 04}

\newpage
\pagestyle{plain}
\setcounter{footnote}{0}

\section{Introduction}

Impurity problems arise in different areas of quantum field theory
and are essential for understandig a number of phenomena
in condensed matter physics. At the experimental side,
the recent interest in point-like impurities (defects) is
triggered by the great progress in building nanoscale devises.

The interaction of quantum fields with impurities represents in general a
hard and yet unsolved problem, but there are relevant achievements
\cite{Delfino:1994nr}--\cite{varna03} in the case of
integable systems in 1+1 space--time
dimensions. The study \cite{Cherednik:vs}--\cite{Liguori:1998xr}
of the special case of purely reflecting impurities (boundaries)
indicates factorized scattering theory
\cite{Yang:1967bm}--\cite{Liguori:de} as the most efficient
method for dealing with this kind of problems. The method provides
on--shell information about
the system and allows to derive the exact
scattering matrix. The goal of the present paper is to extend this framework,
exploring the possibility to recover off-shell information and to reconstruct
the quantum fields, generating the above scattering matrix.
We test this possibility on one of the most extensively studied
integrable systems --
the non--linear Schr\"odinger (NLS) model
\cite{Sklyanin:ye}--\cite{Gattobigio:1998si}.
More precisely, we are concerned below with the NLS model coupled to
a delta--function impurity.
The basic tool of our investigation is a specific exchange algebra
\cite{Mintchev:2002zd, Mintchev:2003ue},
called reflection--transmission (RT) algebra. The RT algebra
is a generalization of the Zamolodchikov--Faddeev (ZF)
\cite{Zamolodchikov:xm, Faddeev:zy}
algebra used in the case without defects. The RT algebra
is originally designed for the construction of the total scattering
operator from the fundamental scattering data, namely the two-body
bulk scattering matrix and the reflection and transmission
amplitudes of a single particle interacting with the defect.
In what follows we demonstrate that in the NLS model
the same algebra allows to reconstruct the
corresponding off-shell quantum field as well. Being the first
exactly solvable example
with non--trivial bulk scattering matrix, the NLS model sheds some light
on the interplay between point--like impurities,
integrability and symmetries. In this respect our solution clarifies
a debated question about the Galilean invariance of the bulk scattering
matrix.

After introducing the model in section \ref{sect1}, we establish the
solution, both at the classical (section \ref{classol}) and second--quantized
(section \ref{quantsol}) levels. We do this in detail, clarifying the
basic properties of the solution. In section \ref{scat} we derive from
the off--shell quantum field the total scattering matrix of the model,
showing that it coincides with the one obtained directly from
factorized scattering.
In section \ref{sect5} we indicate some generalizations.  Our
conclusions and ideas about
further developments are also collected there. The appendices A--C
are devoted to the
proofs of some technical results.

We present below the analysis of the so-called $\delta$--type impurity.
A wider class of defects, interacting with the NLS model and preserving
its integrability,  can be treated in a similar way \cite{NLSlet}. We
have chosen
to focus here on the particular $\delta$--type
defect in order to keep the length of the proofs reasonable, refering to
\cite{NLSlet} for a more physically--oriented treatment of the
general case (without detailed proofs).

\section{Introducing an impurity in the NLS
model\label{sect1}}

We start by recalling some well-known results about the NLS model
without impurity. The reason for this is
twofold: first, because this is a good guide to tackle the problem with
impurity and second, because the central piece of the solution
of the NLS model, the Rosales expansion \cite{Ros, Fokas:1995rj},
can be adapted to the impurity case.

\subsection{The model to solve}
The field theoretic version of NLS is described by a classical
complex field $\Phi(t,x)$ whose equation of motion reads
\begin{equation}
\label{classical-NLS}
(i\partial_t+\partial_x^2)\Phi(t,x)=2g|\Phi(t,x)|^2\Phi(t,x)\, .
\end{equation}
The corresponding action takes the form
\begin{equation}
\ca_{NLS}=\int_{\RR}dt\int_{\RR}dx\left(i\bPhi(t,x)
\partial_t\Phi(t,x)-|\partial_x\Phi(t,x)|^2-g
|\Phi(t,x)|^4\right)\, ,
\end{equation}
and, being in particular invariant under time translation, ensures
the conservation of the energy
\begin{equation}
\ce_{NLS}=\int_{\RR}dx\left(|\partial_x\Phi(t,x)|^2+g
|\Phi(t,x|^4\right)~.
\end{equation}
The latter is non-negative for $g\ge 0$.

It is well-known that this is a nonrelativistic
integrable model \cite{Zak} (see also \cite{Gut} for a review) and an
explicit solution for
the field was given by Rosales in \cite{Ros}
\begin{equation}
\label{solution-rosales1}
\Phi(x,t)=\sum_{n=0}^{\infty}(-g)^n\Phi^{(n)}(x,t)\, ,
\end{equation}
where
\begin{eqnarray}
\label{solution-rosales2}
\Phi^{(n)}(x,t)=\int_{\RR^{2n+1}}\prod_{\topa{i=1}{j=0}}^n
\frac{dp_i}{2\pi}\frac{dq_j}{2\pi}\,
\blambda(p_1)\ldots\blambda(p_n)\lambda(q_n)\ldots\lambda(q_0)
\frac{e^{i\sum\limits_{j=0}^n(q_j x-q^2_j t)-i
\sum\limits_{i=1}^n(p_i x-p^2_i
t)}}{\prod\limits_{i=1}^n(p_i-q_{i-1})(p_i-q_i)}
\end{eqnarray}
and $~\bar{}~$ denotes complex conjugation.

The level $n=0$ is the linear part of the field corresponding to
the free Schr\"odinger equation. It was argued in \cite{Gattobigio:1998si} that
this solution is well-defined for a large class of
functions $\lambda$ (containing the Schwarz space $\cs(\RR)$) and
an upper bound for $g$ was given for the series
(\ref{solution-rosales1}) to converge uniformly in $x$. It also
represents a physical field since it vanishes as $x\to\pm\infty$.
In the same paper, the authors considered NLS on the half-line
$\RR^+$, which can be seen as the model on the whole line in the
presence of a purely reflecting impurity sitting at the origin.
Therefore, the latter represents a particular case of the model
with transmitting and reflecting impurity at $x=0$ we wish to
contemplate in this article. They gave the following action
\begin{equation*}
\ca_{R}=\int_{\RR}dt\int_{\RR^+}dx\left(i\bPhi(t,x)
\partial_t\Phi(t,x)-|\partial_x\Phi(t,x)|^2-g
|\Phi(t,x)|^4\right)-\eta\int_{\RR}dt|\Phi(t,0)|^2\, ,
\end{equation*}
where $\eta\in\RR$ is the parameter controlling the boundary
condition
\begin{equation}
\lim_{x\to 0^+}(\partial_x-\eta)\Phi(t,x)=0\, .
\end{equation}
In our case, since the impurity is allowed to reflect and
transmit, we have to take the $\RR^-$ part into account and we are
led to work with the following action
\begin{equation}
\ca_{RT}=\ca_+ +\ca_- +\ca_{0}\, ,
\end{equation}
where
\begin{eqnarray}
\ca_\pm&=&\int_{\RR}dt\int_{\RR^\pm}dx\left(i\bPhi(t,x)
\partial_t\Phi(t,x)-|\partial_x\Phi(t,x)|^2-g
|\Phi(t,x)|^4\right)\, ,\\
\ca_0&=&-2\eta\int_{\RR}dt|\Phi(t,0)|^2\, .
\end{eqnarray}
The form of $\ca_{RT}$ shows the particular status of the origin
$x=0$ where the impurity sits. Again, the invariance of the action
under time translations ensures the conservation of the energy
\begin{equation}
\ce_{RT}=\int_{\RR^-\oplus\RR^+}dx\Big(|\partial_x\Phi(t,x)|^2+g
|\Phi(t,x)|^4\Big)+2\eta |\Phi(t,0)|^2~.
\end{equation}
It is positive for $g\ge 0,\eta\ge 0$, which is what we assume in
the rest of this article. We will see that $\eta$ characterizes
the transmission and reflection properties of the impurity. Using
the variational principle, one deduces the equation of motion and
the boundary conditions for the field: $\Phi(t,x)$ must be solution of
NLS on $\RR^-$ and $\RR^+$, continuous at $x=0$ and satisfy a
"jump condition" at the origin. It must also vanish at infinity
as a physical field.
\begin{defi}
\label{NLSI}
The nonlinear Schr\"odinger model with a transmitting
and reflecting impurity at the origin is described by the
following boundary problem for the field $\Phi(t,x)$
\begin{eqnarray}
\label{NLS-0}
(i\partial_t+\partial_x^2)\Phi(t,x)-2g|\Phi(t,x)|^2\Phi(t,x)&=&0,
~~x\neq
0\, ,\\
\label{continu}
\lim_{x\to 0^+}\{\Phi(t,x)-\Phi(t,-x)\}&=&0\, ,\\
\label{saut} \lim_{x\to
0^+}\{(\partial_x\Phi)(t,x)-(\partial_x\Phi)(t,-x)
\}-2\eta~\Phi(t,0)&=&0\\
\label{vanish-classique}
\lim_{x\to \pm\infty}\Phi(t,x)&=&0
\end{eqnarray}
\end{defi}

\subsection{Explicit solution\label{classol}}

As announced, the Rosales solution \cite{Ros} can be adapted
suitably to solve the problem of definition \ref{NLSI}. Since
(\ref{solution-rosales1}) is a solution of NLS on $\RR$, it is
easy to devise a solution for (\ref{NLS-0}). Starting from two
copies of (\ref{solution-rosales1}) and (\ref{solution-rosales2}),
one based on a function $\lambda_+$ and the other on a function
$\lambda_-$, denoted $\Phi_+(t,x)$ and $\Phi_-(t,x)$ respectively,
we define
\begin{equation}
\label{solution-NLS-0}
\Phi(t,x)=
\begin{cases}
\Phi_+(t,x)&,~x>0~,\\
\Phi_-(t,x)&,~x<0~,\\
\frac{1}{2}\left(\Phi_+(t,0)+\Phi_-(t,0) \right)&,~x=0
\end{cases}
\end{equation}
It is clearly solution of (\ref{NLS-0}) for $x\neq0$ and from the
vanishing of $\Phi_\pm(t,x)$ as $x\to\pm\infty$,
(\ref{vanish-classique}) is also satisfied. However, there is no
reason why, in general, $\Phi(t,x)$ so defined should satisfy the
boundary conditions (\ref{continu})-(\ref{saut}). In order to
satisfy these conditions, we parametrize $\lda_+, \lda_-$ as
follows
\begin{equation}
\left(\begin{array}{cc} \lambda_+(p) \\ \lambda_-(p)
\end{array}\right) =
\left(\begin{array}{cc} 1 & T(p)\\
T(-p)&1 \end{array}\right)
\left(\begin{array}{cc} \mu_+(p) \\
\mu_-(p)\end{array}\right) +
\left(\begin{array}{cc} R(p) & 0\\
0&R(-p) \end{array}\right)
\left(\begin{array}{cc} \mu_+(-p) \\
\mu_-(-p)\end{array}\right)\, ,
\end{equation}
where
\begin{equation}
\label{coef-TR}
T(p)=\frac{p}{p+i\eta}~,\qquad R(p)=\frac{-i\eta}{p+i\eta}~~,~~p\in\RR~.
\end{equation}
and $\mu_{\pm}(p)$ are arbitrary Schwarz test functions. Then, the
functions $\lda_{\pm}(p)$ satisfy
\begin{equation}
\label{relation-RT}
\lda_\pm(p)=T(\pm p)\lda_{\mp}(p)+R(\pm p)\lda_\pm(-p)\,,\quad \forall
p\in\RR
\end{equation}
which follows from the identities
\begin{equation}
R(p)R(-p)+T(p)T(-p)=1\mb{and} T(p)R(-p)+R(p)T(-p)=0\,,\quad \forall
p\in\RR\,.
\end{equation}
These relations plus a particular choice for the form of $\mu_\pm$
will be essential in the proof of the theorem
\ref{boundary-conditions} below.

 Anticipating the quantum case, if
we interpret $\lda_+$ (resp. $\lda_-$) as a wave-packet,
(\ref{relation-RT}) shows that each wave-packet in $\RR^+$ (resp.
$\RR^-$) is equivalent to the superimposition of a transmitting
part coming from $\RR^-$ (resp. $\RR^+$) and a reflected part in
$\RR_{+}$ (resp. $\RR_{-}$). This physical interpretation will
show up in the next section when we construct a Fock
representation of the creation and annihilation operators.

We are now in position to state the main result
of this section whose lengthy proof we defer until appendix
\ref{app-2}.
\begin{theo}
\label{boundary-conditions} Let $\mu_+$, $\mu_-$ be given by
\begin{equation}
\label{form_mu}
\mu_\pm(k)=\pm\frac{\mu_0(\pm k)+(k\mp
i\eta)~\mu_1(k)}{k\mp i\eta+1}
\end{equation}
where $\mu_0,\mu_1$ are arbitrary Schwartz functions, $\mu_1$
being even and let $\Phi_+(t,x)$, $\Phi_-(t,x)$ be given by the
Rosales expansion
(\ref{solution-rosales1})-(\ref{solution-rosales2}) with $\lda$
replaced by $\lda_+$ and $\lda_-$ respectively. Then, $\Phi(t,x)$
as defined in (\ref{solution-NLS-0}) satisfies the boundary
conditions (\ref{continu}) and (\ref{saut}), \ie
\begin{eqnarray*}
&\displaystyle\lim_{x\to 0^+}&\{\Phi(t,x)-\Phi(t,-x)\}=0\, ,\\
&\displaystyle\lim_{x\to
0^+}&\{(\partial_x\Phi)(t,x)-(\partial_x\Phi)(t,-x)
\}-2\eta\,\Phi(t,0)=0
\end{eqnarray*}
\end{theo}
With this result, we can say that $\Phi(t,x)$ rewritten as
\begin{equation}
\Phi(t,x)=\theta(x)\Phi_+(t,x)+\theta(-x)\Phi_-(t,x)
\end{equation}
where $\theta(x)$ is the Heaviside function defined here to be
$\frac{1}{2}$ at $x=0$, is the classical solution of the nonlinear
Schr\"odinger model with impurity as given in definition
\ref{NLSI}.

We want to emphasize that these boundary conditions decouple for
the nonlinear part of the field (as shown in appendix \ref{app-2})
and this is due to the reflection-transmission property
(\ref{relation-RT}) satisfied by $\lda_+$ and $\lda_-$. This
already gives a good hint that the construction of a local field
from the quantum counterparts of $\lda_+$, $\lda_-$ is achievable,
as we now explain.

\section{Quantization of the system\label{quantsol}}

In this section, we move on to the construction and resolution of
the quantized version of NLS with impurity. As we mentioned
earlier on, the crucial ingredient is the RT algebra which encodes
the properties of the impurity.

\subsection{Reflection-Transmission algebra}
Here we rely on the constructions developed in \cite{Mintchev:2003ue} and
recast them in the particular context of the scalar nonlinear
Schr\"odinger model (no internal degrees of freedom, special form
of the exchange matrix and of the generators, see also
\cite{varna03}).

We consider the associative algebra with identity element $\bf 1$
and two sets of generators,
$\{a_{\alpha}(p),\adag_\alpha(p);~p\in\RR,~\alpha=\pm\}$ and
$\{r(p),t(p);~p\in\RR\}$, called the bulk and defect (reflection
and transmission) generators. The label $\alpha=\pm$ refers to the
half-line $\RR^{\pm}$ with respect to the impurity (in practice it
will indicate where the particle is created or annihilated).
Introducing the measurable function $S:\RR\times\RR\rightarrow\CC$
defined by
\begin{equation}
S(p)=\frac{p-ig}{p+ig}
\end{equation}
the $\cs$-matrix is defined in our context by
\begin{equation}
\cs=\sum_{\alpha_1,\alpha_2=\pm}\cs_{\alpha_1\alpha_2}(p_1,p_2)
E_{\alpha_1\alpha_1}\otimes
E_{\alpha_2\alpha_2}\, ,
\label{bulks}
\end{equation}
where $\cs_{\alpha_1\alpha_2}(p_1,p_2)=S(\alpha_1 p_1-\alpha_2
p_2)$ and
$(E_{\alpha\beta})_{_{\sigma\gamma}}=\delta_{\alpha\sigma}
\delta_{\beta\gamma}$.
It is easy to check that $\cs$ satisfies the unitarity condition
and the quantum Yang-Baxter equation
\begin{equation}
\cs_{12} (p_1,p_2)\, \cs_{21}(p_2,p_1) = \1\otimes \1 \, ,
\end{equation}
\begin{equation}
\cs_{12}(p_1,p_2) \cs_{13}(p_1,p_3) \cs_{23} (p_2,p_3) =
\cs_{23}(p_2,p_3) \cs_{13}(p_1,p_3) \cs_{12}(p_1,p_2) \, .
\end{equation}
Our defect generators $r(p),t(p)$ are related to
$r_\alpha^\beta(p),t_\alpha^\beta(p)$ defined in \cite{Mintchev:2003ue} by
\begin{equation}
r_\alpha^\beta(p)=\delta_\alpha^\beta\, r(\alpha p)
\quad\mbox{and}\quad t_\alpha^\beta(p)=\epsilon_\alpha^\beta\,
t(\alpha p) \quad\mbox{with}\quad \epsilon=
\left(\begin{array}{cc}
0&1\\1&0
\end{array}\right).
\end{equation}
All this setup gives rise to a particular RT algebra whose
defining relations then read
\begin{itemize}
\item Bulk exchange relations
\begin{eqnarray}
\label{bulk-algebra1}
&&a_{\alpha_1}(p_1)~a_{\alpha_2}(p_2)-
S(\alpha_2p_2-\alpha_1p_1)~a_{\alpha_2}(p_2)~a_{\alpha_1}(p_1)=0\\
\label{bulk-algebra2}
&&\adag_{\alpha_1}(p_1)~\adag_{\alpha_2}(p_2)-
S(\alpha_2p_2-\alpha_1p_1)~
\adag_{\alpha_2}(p_2)~\adag_{\alpha_1}(p_1)
=0\\
\label{bulk-algebra3}
&&a_{\alpha_1}(p_1)~\adag_{\alpha_2}(p_2)-
S(\alpha_1p_1-\alpha_2p_2)~\adag_{\alpha_2}(p_2)~
a_{\alpha_1}(p_1) = \nonu
&&\qquad
2\pi\delta(p_1-p_2)~\left[\delta_{\alpha_1}^{\alpha_2}{\bf 1}
+\epsilon_{\alpha_1}^{\alpha_2}~t(\alpha_1p_1)\right]
+2\pi\delta(p_1+p_2)~\delta_{\alpha_1}^{\alpha_2}~r(\alpha_1p_1)~;
\end{eqnarray}
\item Defect exchange relations
\begin{eqnarray}
\label{defect-algebra1} {\left[r(p_1)\, , r(p_2)\right]}
&=& 0\,  ; \\
\label{defect-algebra2} {\left[t(p_1)\, , t(p_2)\right]}
&=& 0\,  ; \\
\label{defect-algebra3} {\left[t(p_1)\, , r(p_2)\right]} &=& 0\,;
\end{eqnarray}
\item Mixed exchange relations
\begin{eqnarray}
\label{mixed-algebra1} a_{\alpha_1}(p_1)\, r(p_2) =
S(p_2-p_1)\,S(p_2+p_1)\, r(p_2)\, a_{\alpha_1}(p_1) \, , \\
\label{mixed-algebra2} r(p_1)\, \adag_{\alpha_2}(p_2)=
S(p_1-p_2)\,S(p_1+p_2)\, \adag_{\alpha_2}(p_2)\, r(p_1)
  \, , \\
\label{mixed-algebra3} a_{\alpha_1}(p_1)\, t(p_2) = S(p_2-p_1)\,
S(p_2+p_1)\,t(p_2)\, a_{\alpha_1}(p_1) \, , \\
\label{mixed-algebra4} t(p_1)\, \adag_{\alpha_2}(p_2) =
S(p_1-p_2)\,S(p_1+p_2)\,\adag_{\alpha_2}(p_2)\,
  t(p_1)\, .
\end{eqnarray}
\item Finally, the defect generators are required to satisfy
unitarity conditions
\begin{eqnarray}
t(p ) t(-p ) + r(p ) r(-p ) = \bf 1 \, , \\
t(p ) r(-p ) + r(p ) t(-p ) = 0 \, ,
\end{eqnarray}
which amount to implement the physical energy
conservation when reflection and transmission occur.
\end{itemize}
Since we aim at second quantize a physical system, we now turn to
the Fock representation of this algebraic setup as it is presented
in \cite{Mintchev:2003ue}. What we need is to represent the generators
$\{a_{\alpha}(p),\adag_\alpha(p),r(p),t(p),~p\in\RR\}$ as
operator-valued distributions acting on a common invariant
subspace of a Hilbert space, $\cf$, to be defined. We should also
identify a normalizable vacuum state $\Omega$ annihilated by $a_{\alpha}$
and cyclic with respect to $\adag_\alpha$. Applying the general
construction of \cite{Mintchev:2003ue}, we know that each such Fock
representation is characterized by two numerical matrices $\ct(p)$
and $\cR(p)$. Here we take
\begin{equation}
\label{forme-TR}
\ct(p)=\left(\begin{array}{cc} 0 &T(p)\\T(-p)&0
\end{array}\right),~~
\cR(p)=\left(\begin{array}{cc} R(p)&0\\0&R(-p)
\end{array}\right)
\end{equation}
with $T,R$ given in (\ref{coef-TR}). Now consider
\begin{equation}
\cl=\bigoplus_{\alpha=\pm}L^2(\RR)
\end{equation}
endowed with the usual scalar product
\begin{equation}
\langle\varphi,\psi\rangle=\int_{\RR}dp~\sum_{\alpha=\pm}
\bvarphi_\alpha(p)\psi_{\alpha}(p)\,,
\end{equation}
which makes it a Hilbert space for the associated norm denoted
$||\cdot ||$. Then, the $n$-particle subspace $\ch^{(n)}$ is the
subspace of the $n$-fold tensor product $\cl^{\otimes n}$ defined
as follows. If $\varphi^{(n)}\in\cl^{\otimes n}$, we identify it
with the column whose entries are
$\varphi^{(n)}_{\alpha_1,\ldots,\alpha_n}$. Then explicitly,
$\ch^{(0)}=\CC$ and for $n\ge 1$, $\varphi^{(n)}\in\ch^{(n)}$ if
and only if
\begin{itemize}
\item  $\varphi^{(n)}\in\cl^{\otimes n}$\, ,
\item
$\varphi^{(n)}_{\alpha_1\ldots\alpha_n}(p_1,\ldots,p_n)=T(\alpha_n
p_n)
\varphi^{(n)}_{\alpha_1\ldots\alpha_{n-1},-\alpha_n}
(p_1,\ldots,p_{n-1},p_n)+$
\begin{equation}
\label{RT-prop} R(\alpha_n p_n)
\varphi^{(n)}_{\alpha_1\ldots\alpha_{n-1}\alpha_n}
(p_1,\ldots,p_{n-1},-p_n)\,,
\end{equation}
\item $n>1$,
$\varphi^{(n)}_{\alpha_1\ldots\alpha_i\alpha_{i+1}\ldots\alpha_n}
(p_1,\ldots,p_i,p_{i+1},\ldots,p_n)=$
\begin{equation}
\label{S-prop}
S(\alpha_ip_i-\alpha_{i+1}p_{i+1})
\varphi^{(n)}_{\alpha_1\ldots\alpha_{i+1}\alpha_i\ldots\alpha_n}
(p_1,\ldots,p_{i+1},p_i,\ldots,p_n)\, ,~1<i<n-1.
\end{equation}
\end{itemize}
The Fock space is
$\displaystyle\cf=\bigoplus_{n=0}^{\infty}\ch^{(n)}$ and the
common invariant subspace is the finite particle space $\cd$
spanned by the linear combination of sequences
$\varphi=(\varphi^{(0)},\varphi^{(1)},\ldots,\varphi^{(n)},\ldots)$
with $\varphi^{(n)}\in\ch^{(n)}$ and $\varphi^{(n)}=0$ for $n$
large enough. $\cd$ is dense in $\cf$. We extend the scalar
product, again denoted by $\langle \cdot\, ,\, \cdot \rangle$, to $\cf$
\begin{eqnarray*}
\forall~\varphi,\psi\in\cf,~~\langle\varphi,\psi\rangle
&=&\sum_{n=0}^\infty\langle\varphi^{(n)},\psi^{(n)}\rangle\nonu
&=&\sum_{n=0}^\infty\int_{\RR^n}dp_1\ldots
dp_n\sum_{\alpha_1,\ldots,\alpha_n=\pm}
~\overline{\varphi}_{\alpha_1\ldots\alpha_n}(p_1,\ldots,p_n)
\psi_{\alpha_1\ldots\alpha_n}(p_1,\ldots,p_n) \, .
\end{eqnarray*}
The unit norm vacuum state is
$\Omega=(1,0,\ldots,0,\ldots)$ and belongs to $\cd$.\\
Now, we can define the action of the smeared bulk operators
$\displaystyle\{a(\mf),\adag(\mf);~\mf\in\oplus_{\alpha=\pm}\cinf\}$
on $\cd$ as follows
\begin{equation}
\label{a-vacuum}
a(\mf) \Omega = 0\, ,
\end{equation}
and for any $\varphi^{(n)}\in\ch^{(n)}$,
\begin{equation}
\left [a(\mf) \varphi \right ]_{\alpha_1\cdots
\alpha_{n-1}}^{(n-1)}(p_1,\ldots,p_{n-1}) = \sqrt{n}
\int_{-\infty}^\infty \frac{dp}{2\pi}~
\sum_{\alpha=\pm}\overline{\mf}_\alpha(p) \varphi_{\alpha \alpha_1
\cdots \alpha_{n-1}}^{(n)} (p, p_1,\ldots,p_{n-1}) \, ,
\end{equation}
\begin{equation}
\label{adag}
\left [\adag(\mf) \varphi \right ]_{\alpha_1\cdots
\alpha_{n+1}}^{(n+1)}(p_1,...,p_{n+1}) = \sqrt {n+1} \left
[P^{(n+1)} \mf\otimes \varphi^{(n)} \right ]_{\alpha_1\cdots
\alpha_{n+1}}(p_1,\ldots,p_{n+1}) \, ,
\end{equation}
where $P^{(n)}$ is the orthogonal projector in $\cl^{\otimes n}$
defined in \cite{Mintchev:2003ue}. For completeness, the explicit form of
(\ref{adag}) is given in appendix \ref{app-3}. These operators are
bounded on each $\ch^{(n)}$
\begin{equation}
\label{estime} \forall~\varphi\in\ch^{(n)},~||a(\mf) \varphi||\le
\sqrt{n}~||\mf||~ ||\varphi||,~~||\adag(\mf) \varphi||\le
\sqrt{n+1}~||\mf||~ ||\varphi||~.
\end{equation}
In particular, they are continuous in the smearing function $\mf$.
Finally, they satisfy
\begin{equation}
\label{a-adag}
\forall~\varphi,\psi\in\cd,~\langle\varphi,a(\mf)\psi\rangle
=\langle\adag(\mf)\varphi,\psi\rangle
~.
\end{equation}
The defect generators are represented as multiplicative operators
on $\cd$, preserving the bulk particle number
\begin{eqnarray}
\left[r(p)\varphi\right]^{(n)}_{\alpha_1\ldots\alpha_n}
(p_1,\ldots,p_n)&=&S(p-\alpha_1
p_1)\ldots S( p-\alpha_n p_n) ~R(p)\nonu
&&\times S(\alpha_n
p_n+p)\ldots
S(\alpha_1p_1+p)\varphi^{(n)}_{\alpha_1\ldots\alpha_n}
(p_1,\ldots,p_n)~,\quad
\end{eqnarray}
\begin{eqnarray}
\left[t(p)\varphi\right]^{(n)}_{\alpha_1\ldots\alpha_n}
(p_1,\ldots,p_n)&=&S(p-\alpha_1
p_1)\ldots S(p-\alpha_n p_n) ~T(p)\nonu
&&\times S(\alpha_n
p_n+p)\ldots S(\alpha_1
p_1+p)\varphi^{(n)}_{\alpha_1\ldots\alpha_n}(p_1,\ldots,p_n)~.\quad
\end{eqnarray}
It follows then that $r$ and $t$ have non-vanishing vacuum
expectation values
\begin{equation}
\langle\Omega,r(p)\Omega\rangle=R(p),~~
\langle\Omega,t(p)\Omega\rangle=T(p)~.
\end{equation}
Introducing finally the operator-valued distributions
$a_\alpha(p),~\adag_\alpha(p)$ as
\begin{equation}
a(\mf)=\int_{\RR}\frac{dp}{2\pi}~\sum_{\alpha=\pm}
\overline{\mf}_\alpha(p)a_\alpha(p),~~
\adag(\mf)=\int_{\RR}\frac{dp}{2\pi}~\sum_{\alpha
=\pm}a^{\dagger}_{\alpha}(p)\mf_\alpha(p)
\end{equation}
one can check that the defining relations of the RT algebra are
satisfied on $\cd$. The operators $a,\adag$ will be referred to as
annihilation and creation operators respectively. Implementing the
automorphism $\varrho$ defined in \cite{Mintchev:2003ue} for which we know
that it is realized by the identity operator for any Fock
representation, we get the quantum analog of the
reflection-transmission property (\ref{relation-RT})
\begin{eqnarray}
a_\alpha(p)&=&\epsilon_\alpha^\beta~ t(\alpha p)~a_\beta(p)+
\delta_\alpha^\beta~ r(\alpha p)~a_\beta(-p)\\
\adag_\alpha(p)&=&\epsilon^\alpha_\beta~\adag_\beta(p)~ t(\beta
p)+\delta^\alpha_\beta~\adag_\beta(-p)~ r(-\beta p)
\end{eqnarray}

\subsection{The question of operator domains}

{F}rom the above it appears that the natural domain to start with is
$\cd$. Actually, it is much too big for practical calculations and
we would like to work on a dense subspace of $\cd$ which would
play the role of the standard formal "state space", a basis of
which is usually denoted by $|k_1,...,k_n\rangle,~~k_1>...>k_n$.
As a first step, we define
\begin{equation}
\begin{array}{l}
\cd_0^0=\CC \\
\cd_0^n=\lbrace\adag_{\alpha_1}(\mf_1)\ldots
\adag_{\alpha_n}(\mf_n)\Omega~
;~\mf_i\in\cinf,~\alpha_i=\pm,i=1,\ldots,n \rbrace,~n\ge 1
\end{array}
\end{equation}
One can check that $\cd_0^n$ is dense in $\ch^{(n)}$, \ie
$\Omega$ is cyclic with respect to $\adag_\alpha$. The
corresponding domain $\cd_0$, dense in $\cd$, is the linear space
of sequences
$\varphi=(\varphi^{(0)},\varphi^{(1)},\ldots,\varphi^{(n)},\ldots)$
with $\varphi^{(n)}\in\cd_0^n$ and $\varphi^{(n)}=0$ for $n$ large
enough. $\cd_0$ is stable under the action of $a_{\alpha}(\mf)$
and $\adag_{\alpha}(\mf)$. Finally, since $T$, $R$ and $S$ are
bounded, $C^\infty$-functions, $\cd_0^n\subset C^\infty_0(\RR^n)$.
Now in order to formulate the desired properties of the quantum
field in the next paragraph, we introduce a partial ordering
relation on $\cinf$ by
\begin{equation}
     \label{ordering}f\succ
g~\Leftrightarrow~\forall x\in supp(f),~\forall y\in
supp(g),~|x|>|y|
\end{equation}
which extends naturally to
$\cinfa,\alpha=\pm$. Let us introduce
\begin{equation}
\label{a-tfourier}
\begin{array}{l}
\widetilde{a}^{\dagger}_{\alpha}(t,x)=\int_{\RR}
~\frac{dp}{2\pi}~\adag_\alpha(p)e^{-ipx+ip^2t},~(t,x)\in\RR^2\\
\null\\
\widetilde{a}^\dagger_\alpha(t,f)=
\int_{\RR}~dx~\widetilde{a}^\dagger_\alpha(t,x)f(x)
,~f\in\cinf
\end{array}
\end{equation}
Now, fix $t\in\RR$ and $\alpha_1,\ldots,\alpha_n$ and define ($vect$
standing for "linear span of")
\begin{equation}
\!\!\!\!\begin{array}{l}
\widetilde{\cd}_0^0=\CC~~\text{and for}~n\ge 1,\\
\widetilde{\cd}_{0,\alpha_1\ldots\alpha_n}^n=vect\lbrace
\widetilde{a}^\dagger_{\alpha_1}(t,f_{1,\alpha_1})
\ldots\widetilde{a}^\dagger_{\alpha_n}(t,f_{n,\alpha_n})
\Omega~;\\
\qquad\qquad\qquad f_{1,\alpha_1}\succ\ldots\succ
f_{n,\alpha_n},~f_{i,\alpha_i}\in
C^\infty_0(\RR^{\alpha_i}),~0\notin
supp(f_{i,\alpha_i}),~i=1,\ldots,n\rbrace
\end{array}
\end{equation}
then the following theorem holds
\begin{theo}
\label{dense}
$\forall~t\in\RR$, $\forall
~\alpha_1,\ldots,\alpha_n=\pm$,
$\widetilde{\cd}_{0,\alpha_1\ldots\alpha_n}^n$ is dense in
$\ch^{(n)}$.
\end{theo}
\prf We only need to consider $n\ge 1$. The proof relies on two
known results of standard analysis. First, the Fourier transform
of a $C^{\infty}$-function with compact support is real analytic
(\ie a Gevrey class 1 function). Second, a real analytic function
vanishing on a given open subset
$U$ of an open connected set $O$, vanishes on the whole of $O$
(see e.g. \cite{MH}).\\
Here, it suffices to show that
$\widetilde{\cd}_{0,\alpha_1\ldots\alpha_n}^n$ is dense in
$\cd_0^n$ for any $t\in\RR$ so let us consider the matrix element
\begin{equation}
\widetilde{A}_{t,\varphi,\alpha_1\ldots\alpha_n}(x_1,\ldots,x_n)
=\langle\varphi^{(n)},
\widetilde{a}^\dagger_{\alpha_1}(t,x_1)\ldots
\widetilde{a}^\dagger_{\alpha_n}(t,x_n)
\Omega\rangle
\end{equation}
where $\varphi^{(n)}\in\cd_0^n$ is arbitrary. To prove the
statement, we now have to show that
\begin{equation}
\label{hypA}
\widetilde{A}_{t,\varphi,\alpha_1\ldots\alpha_n}(x_1,\ldots,x_n)=0,
~~\forall~|x_1|>\ldots>|x_n|>0,~x_i\in\RR^{\alpha_i},~i=1,\ldots,n
\end{equation}
implies $\varphi^{(n)}=0$. From (\ref{a-tfourier}), we get
\begin{equation}
\widetilde{A}_{t,\varphi,\alpha_1\ldots\alpha_n}(x_1,\ldots,x_n)
=\int_{\RR^n}\prod_{j=1}^n~
\frac{dp_j}{2\pi}~e^{-i p_j x_j+i t p_j^2}\langle\varphi^{(n)},
\adag_{\alpha_1}(p_1)\ldots\adag_{\alpha_n}(p_n)\Omega\rangle
\end{equation}
which shows that
$\widetilde{A}_{t,\varphi,\alpha_1\ldots\alpha_n}$ is the Fourier
transform of a $C^\infty$-function with compact support and is
therefore real analytic. Condition (\ref{hypA}) amounts to saying
that $\widetilde{A}_{t,\varphi,\alpha_1\ldots\alpha_n}$ vanishes
on the set
\begin{equation}
U_{\alpha_1\ldots\alpha_n}=\lbrace x\in\RR^n~s.t.~
|x_1|>\ldots>|x_n|>0,~x_i\in\RR^{\alpha_i},~i=1,\ldots,n\rbrace~.
\end{equation} $U_{\alpha_1\ldots\alpha_n}$ being an open subset of
(the open and connected space)
$\RR^n$, we conclude that
$\widetilde{A}_{t,\varphi,\alpha_1\ldots\alpha_n}$ vanishes on
$\RR^n$. This gives in turn that
\begin{equation}
\langle\varphi^{(n)},
\adag_{\alpha_1}(p_1)\ldots\adag_{\alpha_n}(p_n)\Omega\rangle=0,
~~\forall~p_j\in\RR,~j=1,\ldots,n~.
\end{equation}
or, equivalently, from the cyclicity of $\Omega$ with respect to
$\adag$
\begin{equation}
\varphi^{(n)}_{\alpha_1\ldots\alpha_n}(p_1,\ldots,p_n)=0,~~
\forall~p_j\in\RR,~j=1,\ldots,n~.
\end{equation}
Now using the properties (\ref{RT-prop},\ref{S-prop}) satisfied by
$\varphi^{(n)}$, we get
\begin{equation}
\varphi^{(n)}_{\alpha_1\ldots\alpha_n}(p_1,\ldots,p_n)=0~~
\forall~p_j\in\RR,~\forall~\alpha_j=\pm,~j=1,\ldots,n
\end{equation}
that is $\varphi^{(n)}=0$. \finprf This theorem will prove to be
fundamental in the sequel to derive the required properties of the
quantum field operator. Indeed, it will be enough to perform all
calculations only on states in
\begin{equation}
\widetilde{\cd}_{0}^{n,\alpha}=
\widetilde{\cd}_{0,\protect\underbrace{\alpha\ldots\alpha}_n}^n
~\text{with}~\alpha=\pm
\end{equation}
and conclude for the whole domain $\cd$ by a continuity argument.
\begin{lem}
\label{smeared-correl} Let $f_{1,\alpha_1}\succ\ldots\succ
f_{n,\alpha_n}$ and $h_{1,\beta_1}\succ\ldots\succ h_{n,\beta_n}$,
then
\begin{equation}
\label{fourier-correl}
\langle\widetilde{a}^\dagger_{\alpha_1}(t,f_{1,\alpha_1})
\ldots\widetilde{a}^\dagger_{\alpha_n}(t,f_{n,\alpha_n})
\Omega,\widetilde{a}^\dagger_{\beta_1}(t,h_{1,\beta_1})
\ldots\widetilde{a}^\dagger_{\beta_n}(t,h_{n,\beta_n})
\Omega\rangle=\prod_{j=1}^n\delta_{\alpha_j\beta_j}\langle
f_{j,\alpha_j},h_{j,\beta_j}\rangle
\end{equation}
In particular, for
$\varphi\in\widetilde{\cd}_{0,\alpha_1\ldots\alpha_n}^n$
represented as
\begin{equation}
\label{norm}
\varphi=\sum_{\beta\in
B}\widetilde{a}^\dagger_{\alpha_1}(t,f^\beta_{1,\alpha_1})\ldots
\widetilde{a}^\dagger_{\alpha_n}(t,f^\beta_{n,\alpha_n})~,~
f^\beta_{1,\alpha_1}
\succ\ldots\succ f^\beta_{n,\alpha_n},~\forall\beta\in B~,
\end{equation}
where $B$ is a finite set, one has $\displaystyle
||\varphi||=||\sum_{\beta\in
B}f^\beta_{1,\alpha_1}\otimes\ldots\otimes f^\beta_{n,\alpha_n}||$
\end{lem}
\prf To get (\ref{fourier-correl}), one uses an induction on $n$
and combines (\ref{a-tfourier}), (\ref{a-adag}),
(\ref{bulk-algebra3}) and (\ref{a-vacuum}) together with the
support conditions on the smearing functions. Using a contour
integral argument, these support conditions imply that all the
contributions arising from the RT algebra vanish except for the
usual $\delta$- term producing the right-hand side. (\ref{norm})
is a mere consequence of (\ref{fourier-correl}).\finprf

\textbf{Remark:} It is important to realize that the $n$ particle
space $\ch^{(n)}$ is the central piece in this construction
and that, on this space, any operation we have considered (scalar
product, creation operator, Fourier transform) is continuous in
the smearing functions. Since $\cinf$ is dense in $\cs(\RR)$, the
Schwarz space, we can extend the above (especially the definition
of $\cd^n_0$) to smearing functions in $\cs(\RR)$.

\subsection{Quantum field}
We start by defining $\Phi(t,f)$ as
\begin{equation}
\Phi(t,f)=\int_{\RR}~dx~\sum_{\alpha=\pm}\overline{f}_{\alpha}(x)
\Phi_\alpha(t,x),~~
f\in\cc~\qquad\text{where}\qquad
\displaystyle\cc=\bigoplus_{\alpha=\pm}\cinfa\,.
\end{equation}
$f$ is viewed as a column vector $f=\left(\begin{array}{c} f_+
\\ f_-\end{array}\right)$ with $f_\alpha\in\cinfa$ and $0\notin
supp(f_\alpha)$. Following the standard argument of \cite{Davies:gc}, we
replace $\lda_\alpha(p),\overline{\lda}_\alpha(p)$ in the Rosales
expansion of the classical field
(\ref{solution-rosales1})-(\ref{solution-rosales2}) by the
operators $a_\alpha(p),\adag_\alpha(p)$ in order to define
\begin{equation}
\Phi_\alpha(t,x)=\sum_{n=0}^{\infty}(-g)^n\Phi^{(n)}_\alpha(t,x)\,,
\qquad
g>0 \end{equation}
\begin{eqnarray}
\mbox{and}\qquad\qquad   \Phi^{(n)}_\alpha(t,x) &=&
  \int_{\RR^{2n+1}}\prod_{\topa{i=1}{j=0}}^n
\frac{dp_i}{2\pi}\frac{dq_j}{2\pi}\,
\adag_\alpha(p_1)\ldots\adag_\alpha(p_n)a_\alpha(q_n)\ldots
a_\alpha(q_0) \nonu
&&\times\frac{e^{i\sum\limits_{j=0}^n(q_j x-q^2_j t)-i
\sum\limits_{i=1}^n(p_i x-p^2_i t)}}
{\prod\limits_{i=1}^n(p_i-q_{i-1}-i\alpha\varepsilon)
(p_i-q_i-i\alpha\varepsilon)}\quad
\end{eqnarray}
where we used an $i\varepsilon$ prescription depending on
$\alpha=\pm$.\\
We now have several requirements to meet for our quantum theory to
be well-defined. We must give a precise meaning to
$\Phi_\alpha(t,x)$, show that the canonical commutation relations
as well as the boundary conditions (\ref{continu})-(\ref{saut})
hold in a sense we shall make precise and that $\Phi_\alpha(t,x)$
is indeed the
quantum solution we look for.\\
So we start by associating $\Phi_\alpha(t,x)$ with the quadratic
form defined on $\cd\times \cd$ by
\begin{equation}
\label{quadratic}
(\varphi,\psi)\mapsto
\langle\varphi,\Phi_\alpha(t,x)\psi\rangle
\end{equation}
$\cd$ containing only finite particle vectors, it is enough to
investigate $\langle\varphi,\Phi^{(n)}_\alpha(t,x)\psi\rangle$ for
arbitrary $n$.
\begin{prop}
$\forall~n\ge
0,~\forall~\varphi,\psi\in\cd,~~(t,x)\mapsto\langle\varphi,
\Phi^{(n)}_\alpha(t,x)\psi\rangle$
is a $C^\infty$ function.
\end{prop}
\prf The proof is the same as in \cite{Gattobigio:1998si}.\finprf We define the
conjugate $\Phidag_\alpha(t,x)$ again as a quadratic form on
$\cd\times \cd$ by
\begin{equation}
\langle\varphi,\Phidag_\alpha(t,x)\psi\rangle=
\langle\Phi_\alpha(t,x)\varphi,\psi\rangle~.
\end{equation}
It has the same smoothness properties and from (\ref{a-adag}), we
get
\begin{eqnarray}
     \Phi^{\dagger(n)}_\alpha(t,x) &=&
  \int_{\RR^{2n+1}}\prod_{\topa{i=1}{j=0}}^n
\frac{dp_i}{2\pi}\frac{dq_j}{2\pi}\, \adag_\alpha(q_0)\ldots
\adag_\alpha(q_n)a_\alpha(p_n) \ldots a_\alpha(p_1)\nonu
&&\times\frac{e^{-i\sum\limits_{j=0}^n(q_j x-q^2_j t)+i
\sum\limits_{i=1}^n(p_i x-p^2_i t)}}
{\prod\limits_{i=1}^n(p_i-q_{i-1}+i\alpha\varepsilon)
(p_i-q_i+i\alpha\varepsilon)}\quad
\end{eqnarray}
Defining the smeared version
\begin{equation}
\Phidag(t,f)=\int_{\RR}~dx~\sum_{\alpha=\pm}
\Phidag_\alpha(t,x)f_\alpha(x),~~f\in
\cc
\end{equation}
we conclude that $\Phi(t,f)$ and $\Phidag(t,f)$ are understood as
quadratic forms on the domain $\cd$ and are related by
\begin{equation}
\label{conjugate}
\langle\varphi,\Phidag_\alpha(t,f)\psi\rangle=
\langle\Phi_\alpha(t,f)\varphi,\psi\rangle~.
\end{equation}
To get true quantum fields, we need to show that these quadratic
forms give rise to operators on $\cd$. This requires the following
two lemmas.
\begin{lem}
\label{echange-a-phi}
$\forall~\varphi,\psi\in\cd$,
\\[1.2ex]
{(i)} For $h_{1,\alpha}\succ\ldots\succ h_{n,\alpha}$,
\begin{equation}
\begin{array}{l}
\langle\varphi,\Phi_\alpha(t,f_\alpha)
\widetilde{a}^\dagger_\alpha(t,h_{1,\alpha})
\ldots\widetilde{a}^\dagger_\alpha(t,h_{n,\alpha})\Omega\rangle=  \\
\qquad\qquad\sum_{j=1}^n\langle f_\alpha,h_{j,\alpha}\rangle
\langle\varphi,\widetilde{a}^\dagger_\alpha(t,h_{1,\alpha})\ldots
\widehat{\widetilde{a}^\dagger_\alpha(t,h_{j,\alpha})}
\ldots\widetilde{a}^\dagger_\alpha(t,h_{n,\alpha})\Omega\rangle
\end{array}
\end{equation}
where the hatted symbol is omitted.
\\
{(ii)} For $h_\alpha\succ f_\alpha$,
\begin{equation}
\langle\varphi,\Phidag_\alpha(t,f_\alpha)
\widetilde{a}^\dagger_\alpha(t,h_\alpha)\psi\rangle=
\langle\varphi,\widetilde{a}^\dagger_\alpha(t,h_\alpha)
\Phidag_\alpha(t,f_\alpha)\psi\rangle
\end{equation}
\\
{(iii)} For $f_\alpha\succ h_{j,\alpha},~j=1,\ldots,n$,
\begin{equation}
\label{linear-part}
\langle\varphi,\Phidag_\alpha(t,f_\alpha)
\widetilde{a}^\dagger_\alpha(t,h_{1,\alpha})
\ldots\widetilde{a}^\dagger_\alpha(t,h_{n,\alpha})\Omega\rangle=
\langle\varphi,\widetilde{a}^\dagger_\alpha(t,f_\alpha)
\widetilde{a}^\dagger_\alpha(t,h_{1,\alpha})
\ldots\widetilde{a}^\dagger_\alpha(t,h_{n,\alpha})\Omega\rangle
\end{equation}
\end{lem}
\prf One just has to apply the order by order technique developed
in \cite{Davies:gc}. The latter heavily relied on the ZF algebra
satisfied by the creation and annihilation operators. Here, one
must take care in addition of the many contributions of the defect
generators but it is remarkable that the RT algebra satisfied by
the bulk and defect operators leads to the same results (using the
support requirements of the smearing functions and the conditions
$g>0,\eta>0$, all the defect contributions vanish). One realizes
in these manipulations, especially in (\ref{linear-part}), that
the contributions of $\Phi,\Phidag$
on $\widetilde{\cd}_{0}^{n,\alpha}$ are carried by
the zeroth order corresponding to the linear problem (it is the
Fourier transform of $a,\adag$).
\finprf
\begin{lem} Given $\varphi_\alpha \in
\widetilde{\cd}_{0}^{n,\alpha},\psi_\alpha \in
\widetilde{\cd}_{0}^{n+1,\alpha}$ and $f_\alpha\in\cinfa$, the
quadratic form (\ref{quadratic}) satisfies the following
boundedness condition
\begin{equation}
|\langle\varphi_\alpha,\Phi_\alpha(t,f_\alpha)\psi_\alpha\rangle|\le
(n+1)||f_\alpha||~||\varphi_\alpha||~||\psi_\alpha||
\end{equation}
\end{lem}
\prf The proof is similar to that given in \cite{Gattobigio:1998si}
and uses lemmas \ref{smeared-correl} and
\ref{echange-a-phi}-$(i)$. \finprf
{F}rom the Riesz lemma and theorem
\ref{dense}, we conclude that
$\Phi_\alpha(t,f_\alpha):\ch^{(n+1)}\to\ch^{(n)}$ is a bounded
operator for any $n\ge 0$. Thus, it defines an operator on the
common invariant domain $\cd$.  The same holds for
$\Phidag_\alpha(t,f_\alpha):\ch^{(n)}\to\ch^{(n+1)}$ by
(\ref{conjugate}). We can therefore collect our results in the
following theorem
\begin{theo}
\label{phi-operator}
$\Phi(t,f),~\Phidag(t,f):\cd\to\cd$ are
Hermitian conjugate, linear operators and satisfy
\begin{equation}
\Phi(t,f)\Omega=0,~~\Phidag(t,f)\Omega=
\widetilde{a}^\dagger(t,f)\Omega
\end{equation}
\end{theo}
Finally, we will have a \textit{nonrelativistic quantum field} if
we prove the canonical commutation relations for $\Phi,\Phidag$.
\begin{theo}
$\{\Phi(t,f),\Phidag(t,f),~f\in\cc\}$ realize a Fock
representation of the equal time canonical commutation relations
on $\cd$
\begin{equation}
\label{phi-phi}
\left[\Phi(t,f_1),\Phi(t,f_2)\right]=0=\left[\Phidag(t,f_1),
\Phidag(t,f_2)\right]
\end{equation}
\begin{equation}
\label{phi-phidag}
\left[\Phi(t,f_1),\Phidag(t,f_2)\right]=\langle
f_1,f_2\rangle
\end{equation}
\end{theo}
\prf  We know that it suffices to compute the commutators on
$\widetilde{\cd}_{0}^{n,+}$ or $\widetilde{\cd}_{0}^{n,-}$ for
arbitrary $n$ and then extend the results by continuity to
$\ch^{(n)}$ and by linearity to $\cd$. From theorem
\ref{phi-operator}, we get that $(i)-(iii)$ of lemma
\ref{echange-a-phi} hold as operator equalities. Let us start with
the first commutator. It is made out of four parts
\begin{eqnarray}
\left[\Phi(t,f_1),\Phi(t,f_2)\right]=
\left[\Phi_+(t,f_{1,+}),\Phi_+(t,f_{2,+})\right]
+\left[\Phi_+(t,f_{1,+}),\Phi_-(t,f_{2,-})\right]\nonu
+\left[\Phi_-(t,f_{1,-}),\Phi_+(t,f_{2,+})\right]
+\left[\Phi_-(t,f_{1,-}),\Phi_-(t,f_{2,-})\right]
\end{eqnarray}
The first and fourth parts of the right-hand side are easily seen
to be zero from $(i)$ of lemma \ref{echange-a-phi}. One has for
$\alpha=\pm$,
\begin{equation}
\begin{array}{l}
\Phi_\alpha(t,f_{1,\alpha})\Phi_\alpha(t,f_{2,\alpha})
\widetilde{a}^\dagger_\alpha(t,h_{1,\alpha})
\ldots\widetilde{a}^\dagger_\alpha(t,h_{n,\alpha})\Omega=\\
\displaystyle\sum_{j=1}^n \sum_{\topa{k=1}{k\neq j}}^n \langle
f_{2,\alpha},h_{j,\alpha}\rangle\langle
f_{1,\alpha},h_{k,\alpha}\rangle~
\widetilde{a}^\dagger_\alpha(t,h_{1,\alpha})\ldots
\widehat{\widetilde{a}^\dagger_\alpha(t,h_{j,\alpha})} \ldots
\widehat{\widetilde{a}^\dagger_\alpha(t,h_{k,\alpha})}
\ldots\widetilde{a}^\dagger_\alpha(t,h_{n,\alpha})\Omega
\end{array}
\end{equation}
which is symmetric under the exchange of $f_1$ and $f_2$ implying
the vanishing of the commutators. As for the mixed terms, one can
check that
\begin{equation}
\label{mixed}
\Phi_\alpha(t,f_{i,\alpha})
\widetilde{a}^\dagger_{-\alpha}(t,h_{1,-\alpha})
\ldots\widetilde{a}^\dagger_{-\alpha}(t,h_{n,-\alpha})
\Omega=0~,~~i=1,2
\end{equation}
implying the vanishing of the second and third commutators on
$\widetilde{\cd}_{0}^{n,-\alpha}$ and hence on $\cd$. Now the
vanishing of $\left[\Phidag(t,f_1),\Phidag(t,f_2)\right]$ on $\cd$
is obtained by Hermitian conjugation. This proves (\ref{phi-phi}).

(\ref{phi-phidag}) is obtained as follows. Again, we split the
commutator into four parts. Now given a state
in $\widetilde{\cd}_{0}^{n,\alpha}$, we assume
$h_{k,\alpha}\succ f_{2,\alpha}\succ h_{k+1,\alpha}$ for some $k$
and using lemma \ref{echange-a-phi}, we compute for $\alpha=\pm$,
\begin{equation}
\Phi_\alpha(t,f_{1,\alpha})\Phidag_\alpha(t,f_{2,\alpha})
\widetilde{a}^\dagger_\alpha(t,h_{1,\alpha})
\ldots\widetilde{a}^\dagger_\alpha(t,h_{n,\alpha})\Omega=\langle
f_{1,\alpha},f_{2,\alpha}\rangle
\widetilde{a}^\dagger_\alpha(t,h_{1,\alpha})
\ldots\widetilde{a}^\dagger_\alpha(t,h_{n,\alpha})\Omega
+\mathfrak{S}\\
\end{equation}
and
\begin{eqnarray}
\Phidag_\alpha(t,f_{2,\alpha})\Phi_\alpha(t,f_{1,\alpha})
\widetilde{a}^\dagger_\alpha(t,h_{1,\alpha})
\ldots\widetilde{a}^\dagger_\alpha(t,h_{n,\alpha})\Omega=
\mathfrak{S}
\end{eqnarray}
where $\mathfrak{S}$ is
$$\displaystyle\sum_{j=1}^n \langle f_{1,\alpha},h_{j,\alpha}\rangle
\widetilde{a}^\dagger_\alpha(t,h_{1,\alpha})\ldots
\widehat{\widetilde{a}^\dagger_\alpha(t,h_{j,\alpha})}
\ldots\widetilde{a}^\dagger_\alpha(t,h_{k,\alpha})
\widetilde{a}^\dagger_\alpha(t,f_{2,\alpha})
\widetilde{a}^\dagger_\alpha(t,h_{k+1,\alpha})
\ldots\widetilde{a}^\dagger_\alpha(t,h_{n,\alpha})\Omega$$ This
gives
\begin{equation}
\left[\Phi_+(t,f_{1,+}),\Phidag_+(t,f_{2,+})\right]+
\left[\Phi_-(t,f_{1,-}),\Phidag_-(t,f_{2,-})\right]=\langle
f_{1,+},f_{2,+}\rangle+\langle f_{1,-},f_{2,-}\rangle=\langle
f_{1},f_{2}\rangle
\end{equation}
\ie the desired contribution. It is then straightforward using
(\ref{mixed}) to verify that the mixed terms do not contribute
$$
\left[\Phi_+(t,f_{1,+}),\Phidag_-(t,f_{2,-})\right]=
\left[\Phi_-(t,f_{1,-}),\Phidag_+(t,f_{2,+})\right]=0
\qquad\qquad\qquad \rule{5pt}{5pt}
$$
Now we prove that $\Omega$ is cyclic with respect to
$\Phidag$ and that $\Phi(t,x)$
is the solution of the quantum nonlinear Schr\"odinger equation
with impurity. Extending the partial ordering $\succ$ to functions
in $\cc$ as follows
\begin{equation}
\text{For}~f,g\in\cc,~f\succ g \Leftrightarrow f_\alpha \succ
g_\alpha,~\alpha=\pm \, ,
\end{equation}
one can prove the following theorems.
\begin{theo}
The space
\begin{equation}
\ch^{(n)}_0=vect\lbrace\Phidag(t,f_1)\ldots\Phidag(t,f_n)\Omega~
;~f_i\in\cc,~i=1,\ldots,n,~~
f_n\succ\ldots\succ f_1\rbrace
\end{equation}
is dense in $\ch^{(n)}$.
\end{theo}
\prf Let $\varphi^{(n)}\in\ch^{(n)}$ and suppose
$$\langle\varphi^{(n)},\Phidag(t,f_1)\ldots\Phidag(t,f_n)
\Omega\rangle=0,~\forall~f_n\succ\ldots\succ f_1~.$$
Then, it is true in particular for $f_{i,-}=0,~i=1,\ldots,n$ but
in that case, we have
$$\Phidag(t,f_1)\ldots\Phidag(t,f_n)\Omega=
\widetilde{a}^\dagger_+(t,f_{n,+})\ldots
\widetilde{a}^\dagger_+(t,f_{1,+})
\Omega$$ which implies $\varphi^{(n)}=0$ since
$\widetilde{\cd}_{0}^{n,+}$ is dense in $\ch^{(n)}$.
\finprf

\begin{theo}
The quantum field $\Phi$ is solution of the quantum nonlinear
Schr\"odinger equation with impurity, \ie it satisfies
\begin{equation}
(i\partial_t+\partial_x^2)\langle\varphi,\Phi(t,x)\psi\rangle=
2g\langle\varphi,:\Phi\Phidag\Phi:(t,x)\psi\rangle
\end{equation}
  and the following boundary conditions
\begin{eqnarray}
\label{continuity}
&\displaystyle\lim_{x\to0^+}&\langle\varphi,\lbrace\Phi_+(t,x)-
\Phi_-(t,-x)
\rbrace\psi\rangle=0\\
\label{jump} &\displaystyle\lim_{x\to
0^+}&\partial_x\langle\varphi,\lbrace
\Phi_+(t,x)+\Phi_-(t,-x)\rbrace\psi\rangle=2\eta\,\lim_{x\to 0}
\langle\varphi, \Phi(t,x)\psi\rangle\\
\label{vanish} &\displaystyle\lim_{x\to \pm\infty}&\langle\varphi,
\Phi(t,x)\psi\rangle=0
\end{eqnarray}
for any $\varphi,\psi\in\cd$.
\end{theo}
\prf Inspired by the classical case, we split the field as follows
\begin{equation}
\label{split}
\Phi(t,x)=\theta(x)\Phi_+(t,x)+\theta(-x)\Phi_-(t,x)~.
\end{equation}
The main difficulty here is to specify a normal ordering
prescription for the analog of the cubic term. We adopt the
prescription detailed in \cite{Gattobigio:1998si} for the normal ordering
denoted $:...:$ and apply it to $\Phi_\alpha,~\alpha=\pm$. Then
following \cite{Gattobigio:1998si} (theorem 5), one gets that the quantum field
$\Phi_\alpha$ is solution of the nonlinear Schr\"odinger equation
on the half-line $\RR^\alpha$: for all $\varphi,\psi\in\cd$,
\begin{equation}
(i\partial_t+\partial_x^2)\langle\varphi,\Phi_\alpha(t,x)\psi\rangle=
2g\langle\varphi,:\Phi_\alpha\Phidag_\alpha\Phi_\alpha:(t,x)\psi\rangle
\end{equation}
The situation is now similar to the classical case and we have to
check the quantum analog of (\ref{continu})-(\ref{vanish-classique}). The
idea lies again in realizing that eqs.
(\ref{continuity})-(\ref{vanish}) can be cast into a
zeroth-order/linear problem. Following the line of argument of
\cite{Gattobigio:1998si} (theorem 6), one shows that given
$\varphi,\psi\in\cd$,
there exists $\chi\in\ch^{(1)}$ such that $\langle\varphi,
\Phi(t,f)\psi\rangle=\langle\Omega, \Phi(t,f)\chi\rangle$ and
$\chi$ is independent of $f$. This gives in particular
$\langle\varphi, \Phi_\alpha(t,f_\alpha)\psi\rangle=\langle\Omega,
\Phi_\alpha(t,f_\alpha)\chi\rangle,~\alpha=\pm$ and we can compute
\begin{equation}
\langle\varphi,
\Phi_\alpha(t,x)\psi\rangle=\langle\widetilde{a}^\dagger_\alpha(t,x)
\Omega,\chi\rangle=\int_{\RR}~\frac{dp}{2\pi}~
e^{ipx-ip^2t}\chi_\alpha(p)
\end{equation}
Then, eqs. (\ref{continuity}-\ref{jump}) are easily obtained using
the property (\ref{RT-prop}) satisfied by $\chi$. Finally, since
$\chi_\alpha\in L^2(\RR)$, $\langle\varphi,
\Phi_\alpha(t,x)\psi\rangle$ as a function of $x$ is also in
$L^2(\RR)$ and therefore vanishes at infinity. Noting that
$\displaystyle\lim_{x\to \pm\infty}\langle\varphi,
\Phi(t,x)\psi\rangle=\displaystyle\lim_{x\to \pm\infty}
\langle\varphi, \Phi_\pm(t,x)\psi\rangle$, we get
(\ref{vanish}).\finprf We have finally achieved the goal of this
section: we have explicitly constructed off-shell local fields for
the quantum nonlinear Schr\"odinger system on the line in the
presence of a transmitting and reflecting impurity. As mentioned
in \cite{Mintchev:2003ue}, this remained a challenging open problem for which
we brought an answer here. In other words, the quantum inverse
scattering method remains valid in the presence of an impurity
provided that the ZF algebra is replaced by the RT algebra.

\section{Scattering theory\label{scat}}

Scattering theory in the presence of an impurity was studied on
general grounds in \cite{Mintchev:2003ue} by introducing the RT algebra which,
being a generalization of the ZF and boundary algebras, is
believed to prove fundamental also in the study of off-shell
correlations functions and symmetries for 1+1-dimensional
integrable systems with impurity.

In this section, we aim at giving some credit to this in the
context of the nonlinear Schr\"odinger model. Indeed from the
above results, we can get some insight in the correlations
functions of the theory. The correlations functions vanish unless
they involve the same number of $\Phi$ and $\Phidag$ and for a
given $2n$-point function, we need at most the first $(n-1)$
order terms
in the Rosales expansion of the field. This reads
\begin{eqnarray}
\langle\Omega,
\Phi(t_1,x_1)\ldots\Phi(t_n,x_n)\Phidag(t_{n+1},x_{n+1})\ldots
\Phidag(t_{2n},x_{2n})
\Omega\rangle=\nonu
\sum_{\topa{K\le n-1}{L\le
n-1}}g^{K+L}\langle\Omega,\Phi^{(k_1)}(t_1,x_1)\ldots
\Phi^{(k_n)}(t_n,x_n)
\Phi^{\dagger(l_1)}(t_{n+1},x_{n+1})\ldots\Phi^{\dagger(l_n)}
(t_{2n},x_{2n})
\Omega\rangle
\end{eqnarray}
where $\displaystyle K=\sum_{i=1}^n k_i$ and $\displaystyle
L=\sum_{i=1}^n l_i$ and the sum runs over all n-uplets
$(k_1,\ldots,k_n),~(l_1,\ldots,l_n)\in\ZZ_+^n$ such that $K,~L\le
n-1$.

One has for example (with $t_{12} = t_1-t_2$,
$x_{12} = x_1-x_2$ and ${\widetilde x}_{12}
= x_1+x_2$):

\begin{eqnarray}
\langle \Omega, \Phi(t_1,x_1)\Phi^\ast (t_2,x_2)\Omega \rangle =
\int_{-\infty}^{+\infty} \frac{dp}{2\pi} e^{-ip^2t_{12}} \cdot
\qquad \qquad \qquad \nonumber \\
\Big\{\theta (x_1)\theta (x_2)\left [e^{ipx_{12}} +
R(p)e^{ip{\widetilde x}_{12}}\right ]+
\theta (-x_1)\theta (-x_2)\left [e^{ipx_{12}} +
{\overline R}(p)e^{ip{\widetilde x}_{12}}\right] +
\nonumber \\
\theta (x_1)\theta (-x_2) T(p)e^{ipx_{12}} +
\theta (-x_1)\theta (x_2) {\overline T}(p)e^{ipx_{12}}\Big\} \,  .
\qquad \qquad \quad
\label{two-point}
\end{eqnarray}

More importantly, using the Haag-Ruelle approach suitably, we can
relate off-shell and asymptotic theories and, doing so, fill the
gap of our quantum field theory. Indeed, on the one hand, we know
from \cite{Mintchev:2003ue} that the Fock representation of the RT algebra
generates the asymptotic states of a general integrable theory
with impurity with corresponding $S$-matrix. On the other
hand, in this paper we constructed off-shell local time-dependent
fields whose behaviour as $t\to\pm\infty$ we would like to know.

\subsection{Asymptotic theory}

The first step is to characterize wave packets for the free
Schr\"odinger equation which take into account the presence of the
impurity at $x=0$. We adopt the following setup. For
$\mf\in\cinf$, we define
\begin{equation}
\label{packet}
f^t(x)=\int_{\RR}\frac{dp}{2\pi}~\mf(p)~e^{ipx-ip^2t}
\end{equation}
We transpose the partial ordering (\ref{ordering}) to functions of
the variable $p$.
\begin{defi}
\label{definition}
Given $n,m\ge 1$, consider two sets of
functions
\begin{equation}
\mathfrak{H}_n=\lbrace \mh_{i,\alpha_i}\in
C^\infty_0(\RR^{\alpha_i}),~i=1,\ldots,n\rbrace
\mb{and}
\mathfrak{G}_m=\lbrace\mg_{i,\beta_i}\in
C^\infty_0(\RR^{-\beta_i}),~i=1,\ldots,m\rbrace
\end{equation}
where the functions obey the following order prescriptions:
\begin{equation}
\label{ordre} \mh_{1,\alpha_1}\succ\ldots\succ\mh_{n,\alpha_n}~,~~
\mg_{m,\beta_m}\succ\ldots\succ\mg_{1,\beta_1}~.
\end{equation}
\end{defi}
We also define
\begin{equation}
h^\theta_{i,\alpha_i}(x)=\theta(\alpha_i
x)h^t_{i,\alpha_i}(x)~,~~g^\theta_{i,\beta_i}(x)=\theta(\beta_i
x)g^t_{i,\beta_i}(x)\,.
\end{equation}
By construction, $h^\theta_{i,\alpha_i}(x)$ represents
wave-packets in $\RR^{\alpha_i}$ moving away from the impurity
towards $\alpha_i\infty$ while $g^\theta_{i,\beta_i}(x)$
represents wave-packets in $\RR^{\beta_i}$ moving towards the
impurity. One already understands that they will be relevant for
the so-called "out" and "in" states respectively. In fact, this is
the main theorem of this section for which we need some
preliminary results.

{F}rom the previous section, we know the exchange and commutation
properties of $\Phidag$ and $\widetilde{a}^\dagger$ smeared with
ordered functions in the variable $x$. Here, our wave-packets were
constructed from ordered functions in $p$ but we made no
assumption as to their ordering in $x$. Therefore, we have to
include all the possibilities and this requires the use of the
permutation group of $n$ elements $\mS_n$. For
$\sigma\in\mS_n,~\pi\in\mS_m,~n,m\ge 2$, we introduce
\begin{eqnarray}
  \theta_h^\sigma(\alpha_1 x_1,\ldots,\alpha_n x_n ) &=&
\prod_{\topa{i,j=1}{i<j}}^n\theta(\alpha_{\sigma_i}x_{\sigma_i}
-\alpha_{\sigma_j}x_{\sigma_j})\\
  \theta_g^\pi(\beta_1 x_1,\ldots,\beta_m x_m) &=&
  \prod_{\topa{i,j=1}{i>j}}^m\theta(\beta_{\pi_i}x_{\pi_i}
-\beta_{\pi_j}x_{\pi_j})
\end{eqnarray}
satisfying
\begin{eqnarray}
\sum_{\sigma\in\mS_n}\theta_h^\sigma(\alpha_1 x_1,\ldots,\alpha_n
x_n)=1=\sum_{\pi\in\mS_m}\theta_g^\pi(\beta_1 x_1,\ldots,\beta_m
x_m)
\end{eqnarray}

\begin{lem}\label{aux-results}
Given any two sets of functions in $\mathfrak{H}_n$ and
$\mathfrak{G}_m$:
\begin{description}
\item{(i)}
The following limits hold
\begin{equation}
\begin{array}{c}
\displaystyle \lim_{t\to
+\infty}||h^\theta_{1,\alpha_1}\otimes\ldots\otimes
h^\theta_{n,\alpha_n} -h^t_{1,\alpha_1}\otimes\ldots\otimes
h^t_{n,\alpha_n}||=0\\
\displaystyle \lim_{t\to
-\infty}||g^\theta_{1,\beta_1}\otimes\ldots\otimes
g^\theta_{m,\beta_m} -g^t_{1,\beta_1}\otimes\ldots\otimes
g^t_{m,\beta_m}||=0
\end{array}
\end{equation}
\item{(ii)} Let $e_n$ be the identity of $\mathfrak{S}_n$ and let us define
\begin{equation}
\begin{array}{c}
H^\sigma_{\alpha_1\ldots\alpha_n}(x_1,\ldots,x_n)
=h^\theta_{1,\alpha_1}(x_1)
\ldots h^\theta_{n,\alpha_n}(x_n)\theta^\sigma_h(\alpha_1
x_1,\ldots,\alpha_n x_n)\, ,\\
G^\pi_{\beta_1\ldots\beta_m}(x_1,\ldots,x_m)
=g^\theta_{1,\beta_1}(x_1)
\ldots g^\theta_{m,\beta_m}(x_m)\theta^\pi_g(\beta_1
x_1,\ldots,\beta_m x_m)\, .
\end{array}
\end{equation}
Then
\begin{equation}
\lim_{t\to +\infty}||H^\sigma_{\alpha_1\ldots\alpha_n}||=0~~,~~
\lim_{t\to -\infty}||G^\pi_{\beta_1\ldots\beta_m}||=0\qquad
\mbox{for all $\sigma\neq e_n,~\pi\neq e_m$.}
\end{equation}
\item{(iii)} The following estimate is valid for any
$F\in L^2(\RR^{n})$
\begin{equation}
\label{major} ||\int_{\RR^n}dx_1\ldots dx_n~F(x_1,\ldots,x_n)
\widetilde{a}^{\dagger}_{\alpha_1}
(t,x_1)\ldots\widetilde{a}^{\dagger}_{\alpha_n}
(t,x_n)||\le\sqrt{n!}~||F||~.
\end{equation}
\end{description}
\end{lem}
\prf The ideas are the same as those detailed in
\cite{Gattobigio:1998si} from
theorem 7 onwards and rest especially on the use of the weak limit
\begin{equation}
\lim_{t\to\pm\infty}\frac{e^{itk}}{k\pm i\varepsilon}=0~.
\end{equation}
We just stress again that in our case all the above holds thanks
to the use of the RT algebra and by paying careful attention to
the support conditions encoded in (\ref{ordre}). \finprf
We are
now in position to identify the asymptotic behaviour of the
field as $t\to\pm\infty$.
\begin{theo}
\label{strong-lim} The following limits hold in the strong sense in
the Fock space $\cf$
\begin{equation}
\label{lim+}
\lim_{t\to
+\infty}\Phidag(t,h^\theta_{1,\alpha_1})\ldots
\Phidag(t,h^\theta_{n,\alpha_n})\Omega=
\adag_{\alpha_1}(\mh_{1,\alpha_1})\ldots
\adag_{\alpha_n}(\mh_{n,\alpha_n})\Omega
\end{equation}
\begin{equation}
\label{lim-}
\lim_{t\to-\infty}
\Phidag(t,g^\theta_{1,\beta_1})\ldots
\Phidag(t,g^\theta_{m,\beta_m})\Omega=
\adag_{\beta_1}(\mg_{1,\beta_1})\ldots
\adag_{\beta_m}(\mg_{m,\beta_m})\Omega
\end{equation}
\end{theo}
\prf We note first that from (\ref{split}) one gets
$\Phidag(t,h^\theta_{i,\alpha_i})=
\Phidag_{\alpha_i}(t,h^\theta_{i,\alpha_i})$ and
$\Phidag(t,g^\theta_{i,\beta_i})=
\Phidag_{\beta_i}(t,g^\theta_{i,\beta_i})$ so that
\begin{equation}
\Phidag(t,h^\theta_{i,\alpha_i})\Omega=
\widetilde{a}^{\dagger}_{\alpha_i}
(t,h^\theta_{i,\alpha_i})\Omega~\text{ and}~~
\Phidag(t,g^\theta_{i,\beta_i})\Omega=
\widetilde{a}^{\dagger}_{\beta_i}
(t,g^\theta_{i,\beta_i})\Omega~.
\end{equation}
Moreover, for $\mf_\alpha\in C^\infty_0(\RR^\alpha)$, one has
\begin{equation}
\adag_\alpha(\mf_\alpha)=\widetilde{a}^\dagger_\alpha(t,f^t_\alpha)~.
\end{equation}
Collecting all this, theorem \ref{strong-lim} is proved for
$n=m=1$ using $(i)$,$(iii)$ of lemma (\ref{aux-results}):
\begin{equation}
||\Phidag(t,f^\theta_{\alpha})\Omega-
\adag_\alpha(\mf_\alpha)\Omega||=
||\widetilde{a}^{\dagger}_{\alpha} (t,f^\theta_{\alpha})\Omega-
\widetilde{a}^\dagger_\alpha(t,f^t_\alpha)\Omega||\le
||f^\theta_\alpha-f^t_\alpha||~,
\end{equation}
$\mf$ playing the role of $\mh$ or $\mg$. Now we want to compute
the left-hand sides of eqs. (\ref{lim+}) and (\ref{lim-}) for
$n,m\ge 2$. We give details for eq.(\ref{lim+}).
$$
\Phidag(t,h^\theta_{1,\alpha_1})\ldots
\Phidag(t,h^\theta_{n,\alpha_n})\Omega=$$
$$\sum_{\sigma\in\mS_n}\int_{\RR^n}dx_1\ldots
dx_n~H^\sigma_{\alpha_1\ldots\alpha_n}(x_1,\ldots,x_n)
\Phidag_{\alpha_1}(t,x_1)\ldots\Phidag_{\alpha_n}(t,x_n)\Omega= $$
$$\sum_{\sigma\in\mS_n}\int_{\RR^n}dx_1\ldots
dx_n~H^\sigma_{\alpha_1\ldots\alpha_n}(x_1,\ldots,x_n)
\widetilde{a}^\dagger_{\alpha_{\sigma_1}}(t,x_{\alpha_{\sigma_1}})
\ldots
\widetilde{a}^\dagger_{\alpha_{\sigma_n}}(t,x_{\alpha_{\sigma_n}})
\Omega=$$
$$\widetilde{a}^\dagger_{\alpha_1}(t,h^\theta_{1,\alpha_1})\ldots
\widetilde{a}^\dagger_{\alpha_n}(t,h^\theta_{n,\alpha_n})\Omega+$$
\begin{eqnarray}
\sum_{\topa{\sigma\in\mS_n}{\sigma\neq e_n}}\int_{\RR^n}dx_1\ldots
dx_n~H^\sigma_{\alpha_1\ldots\alpha_n}(x_1,\ldots,x_n)\lbrace
\widetilde{a}^\dagger_{\alpha_{\sigma_1}}(t,x_{\alpha_{\sigma_1}})
\ldots
\widetilde{a}^\dagger_{\alpha_{\sigma_n}}(t,x_{\alpha_{\sigma_n}})
\Omega\nonu
-\widetilde{a}^\dagger_{\alpha_1}(t,h^\theta_{1,\alpha_1})\ldots
\widetilde{a}^\dagger_{\alpha_n}(t,h^\theta_{n,\alpha_n})\Omega
\rbrace
\end{eqnarray}
where we used point $(iii)$ of lemma \ref{echange-a-phi} and
(\ref{phi-phi}) for $\Phidag$ in the second equality. Applying
(\ref{major}) then gives
\begin{eqnarray}
||\Phidag(t,h^\theta_{1,\alpha_1})\ldots
\Phidag(t,h^\theta_{n,\alpha_n})\Omega
-\adag_{\alpha_1}(\mh_{1,\alpha_1})\ldots\adag_{\alpha_n}
(\mh_{n,\alpha_n})\Omega
||\le \nonu \sqrt{n!}|| h^\theta_{1,\alpha_1}\otimes\ldots\otimes
h^\theta_{n,\alpha_n} -h^t_{1,\alpha_1}\otimes\ldots\otimes
h^t_{n,\alpha_n}||+2\sqrt{n!}\sum_{\topa{\sigma\in\mS_n}{\sigma\neq
e_n}}||H^\sigma_{\alpha_1\ldots\alpha_n} ||
\end{eqnarray}
implying (\ref{lim+}) by points $(i)$-$(ii)$ of lemma
\ref{aux-results}. Similar computations give
\begin{eqnarray}
||\Phidag(t,g^\theta_{1,\beta_1})\ldots
\Phidag(t,g^\theta_{m,\beta_m})\Omega
-\adag_{\beta_1}(\mg_{1,\beta_1})\ldots
\adag_{\beta_m}(\mg_{m,\beta_m})\Omega
||\le \nonu \sqrt{m!}|| g^\theta_{1,\beta_1}\otimes\ldots\otimes
g^\theta_{m,\beta_m} -g^t_{1,\beta_1}\otimes\ldots\otimes
g^t_{m,\beta_m}||+2\sqrt{m!}\sum_{\topa{\pi\in\mS_m}{\pi\neq
e_m}}||G^\pi_{\beta_1\ldots\beta_m}||
\end{eqnarray}
proving (\ref{lim-}).\finprf

\subsection{Scattering matrix}

Now that we have identified the natural "free" dynamics approached
by our interacting field as $t\to\pm\infty$, we are left with the
verification of asymptotic completeness allowing the construction
of a unitary $S$-matrix. We emphasize here that our "in" and "out"
spaces are slightly different from those exhibited in \cite{Mintchev:2003ue}
because of our ordering involving absolute values, so that we have
to re-check their properties.
\begin{prop}
\label{complete}
Let
\begin{equation}
\cf^{in}=vect\{\Omega,\adag_{\beta_1}(\mg_{1,\beta_1})
\ldots\adag_{\beta_m}(\mg_{m,\beta_m})\Omega,~
\beta_i=\pm,~i=1,\ldots,m,~m\ge 1 \}~,
\end{equation}
\begin{equation}
\cf^{out}=vect\{\Omega,\adag_{\alpha_1}(\mh_{1,\alpha_1})
\ldots\adag_{\alpha_n}(\mh_{n,\alpha_n})\Omega,~
\alpha_i=\pm,~i=1,\ldots,n,~n\ge 1 \}~,
\end{equation}
where $\mh_{i,\alpha_i}$ and $\mg_{j,\beta_j}$ run over
$\mathfrak{H}_n$ and $\mathfrak{G}_m$.\\
Then, $\cf^{in}$ and $\cf^{out}$ are separately dense in $\cf$.
\end{prop}
\prf We deal with $\cf^{in}$. Again, it is sufficient to consider
the matrix element
\begin{equation}
A_{t,\varphi,\beta_1\ldots\beta_p}(p_1,\ldots,p_m)=
\langle\varphi^{(n)},
\adag_{\beta_1}(t,p_1)\ldots\adag_{\beta_m}(t,p_m) \Omega\rangle
\end{equation}
where $\varphi^{(n)}\in\ch^{(n)}$ is arbitrary and to show that
\begin{equation}
\label{A}
A_{t,\varphi,\beta_1\ldots\beta_p}(p_1,\ldots,p_m)=0~,~~\forall~
|p_1|<\ldots<|p_m|,~p_i\in\RR^{-\beta_i},
\beta_i=\pm,~i=1,\ldots,m
\end{equation}
implies $\varphi^{(n)}=0$. From the cyclicity of $\Omega$ with
respect to $\adag$, (\ref{A}) gives
\begin{equation}
\varphi^{(n)}_{\beta_1\ldots\beta_p}(p_1,\ldots,p_m)=0~,~~\forall~
|p_1|<\ldots<|p_m|,~p_i\in\RR^{-\beta_i},
\beta_i=\pm,~i=1,\ldots,m
\end{equation}
and in view of the properties of $\varphi^{(n)}\in\ch^{(n)}$, this
implies in turn
\begin{equation}
\varphi^{(n)}_{\beta_1\ldots\beta_p}(p_1,\ldots,p_m)=0~,~~\forall~
p_i\in\RR,~ \beta_i=\pm,~i=1,\ldots,m
\end{equation}
\ie $\varphi^{(n)}=0$. The case of $\cf^{out}$ is similar. \finprf
We turn to the definition of the scattering operator $\bf S$ of
our theory.
\begin{prop}
Take functions in $\mathfrak{H}_n$ and let ${\bf S}:
\cf^{out}\to\cf^{in}$ act as follows
\begin{eqnarray}
&&\label{scattering-S}
{\bf S}~\Omega=\Omega~~\text{and}~~{\bf
S}:\adag_{\alpha_1}(\mh_{1,\alpha_1})\ldots\adag_{\alpha_n}
(\mh_{n,\alpha_n})\Omega\mapsto
\adag_{\alpha_n}(\widehat{\mh}_{n,\alpha_n})\ldots
\adag_{\alpha_1}(\widehat{\mh}_{1,\alpha_1})\Omega
\qquad \\
&&\mbox{where}\qquad
\label{p-p}
\widehat{\mh}_{i,\alpha_i}(p)=\mh_{i,\alpha_i}(-p)\in\mathfrak{G}_n~.
\end{eqnarray}
Then $\bf S$ is invertible and $\bf S$, $\bf S^{-1}$ are unitary
operators acting on $\cf$.
\end{prop}
\prf From the definitions (\ref{scattering-S}) and (\ref{p-p}),
one deduces immediately that $\bf S^{-1}$ is well-defined.
Then, it is
straightforward, albeit lengthy, to check that
\begin{eqnarray}
\langle{\bf
S}~\adag_{\alpha_1}(\mh_{1,\alpha_1})\ldots
\adag_{\alpha_n}(\mh_{n,\alpha_n})\Omega,{\bf S}~
\adag_{\gamma_1}(\mf_{1,\gamma_1})\ldots
\adag_{\gamma_n}(\mf_{n,\gamma_n})\Omega\rangle=\nonu
\langle\adag_{\alpha_1}(\mh_{1,\alpha_1})\ldots
\adag_{\alpha_n}(\mh_{n,\alpha_n})\Omega,
\adag_{\gamma_1}(\mf_{1,\gamma_1})\ldots
\adag_{\gamma_n}(\mf_{n,\gamma_n})\Omega\rangle~.
\end{eqnarray}
In evaluating the left-hand side, one just has to notice that all
the contributions coming from the defect generators vanish due to
the support properties of the smearing functions and one is left
with what would be obtained by using the ZF algebra. Then, it is
just a matter of changing the variables into their opposite to get
the right-hand side.

Next, following the line of argument given in \cite{Liguori:1998xr},
one
extends $\bf S$ to $\cf^{out}$ by linearity, preserving unitarity.
This gives rise to bounded linear operators which one can uniquely
extend by continuity to the whole of $\cf$. We note that this last
step is allowed by the asymptotic completeness property satisfied
by $\cf^{out}$ and $\cf^{in}$ (cf Proposition \ref{complete}). The
case of $\bf S^{-1}$ is similar.\finprf

Refering now to \cite{Mintchev:2003ue} and we finish the description
of our scattering theory by defining the correspondence between
$in$ and $out$ states and the asymptotic states identified in
theorem \ref{strong-lim} (correspondence already anticipated in
our calling $\cf^{out}$ and $\cf^{in}$ the "in" and "out" spaces).
\begin{eqnarray}
|\mg_{1,\beta_1};\ldots;\mg_{m,\beta_m}\rangle^{in}=
\adag_{\beta_1}(\mg_{1,\beta_1})\ldots
\adag_{\beta_m}(\mg_{m,\beta_m})\Omega\\
|\mh_{1,\alpha_1};\ldots;\mh_{n,\alpha_n}\rangle^{out}=
\adag_{\alpha_1}(\mh_{1,\alpha_1})\ldots
\adag_{\alpha_n}(\mh_{n,\alpha_n})\Omega
\end{eqnarray}
Transition amplitudes are therefore easily computable from
\begin{eqnarray}
^{out}\langle\mh_{1,\alpha_1};\ldots;
\mh_{n,\alpha_n}|\mg_{1,\beta_1};\ldots;\mg_{m,\beta_m}\rangle^{in}
=\nonu
\langle
\adag_{\alpha_1}(\mh_{1,\alpha_1})\ldots
\adag_{\alpha_n}(\mh_{n,\alpha_n})\Omega,
\adag_{\beta_1}(\mg_{1,\beta_1})\ldots
\adag_{\beta_m}(\mg_{m,\beta_m})\Omega\rangle
\end{eqnarray}
and using (\ref{a-adag}), (\ref{bulk-algebra3}),
(\ref{mixed-algebra2}), (\ref{mixed-algebra4}) and
(\ref{a-vacuum}). One recovers for transition amplitudes that they
vanish unless $n=m$ as expected for an integrable system where
particle production does not occur. As an example, we derive in
our context the one and two particle transition amplitudes
obtained in \cite{Mintchev:2002zd}. We start with the computation of the
correlators
\begin{equation}
\langle
\adag_{\alpha}(p)\Omega,\adag_{\beta}(q)
\Omega\rangle=\delta_{\alpha}^{\beta}\delta(p-q)+
\epsilon_{\alpha}^{\beta}\delta(p-q)T(\alpha
p)+\delta_{\alpha}^{\beta}\delta(p+q)R(\alpha p)
\end{equation}
and
\begin{eqnarray}
& &\langle
\adag_{\alpha_1}(p_1)\adag_{\alpha_2}(p_2)
\Omega,\adag_{\beta_1}(q_1)\adag_{\beta_2}(q_2)\Omega\rangle=\nonu
& &S(\alpha_1 p_1-\beta_1 q_1)\left[
\delta_{\alpha_2}^{\beta_1}+\epsilon_{\alpha_2}^{\beta_1}\,T(\alpha_2
p_2)\right]\left[
\delta_{\alpha_1}^{\beta_2}+\epsilon_{\alpha_1}^{\beta_2}\,T(\alpha_1
p_1)\right]\delta(p_2-q_1)\,\delta(p_1-q_2)\nonu &+&S(\alpha_1
p_1-\beta_1 q_1)\left[\delta_{\alpha_2}^{\beta_1}\,R(\alpha_2
p_2)\right]\left[
\delta_{\alpha_1}^{\beta_2}+\epsilon_{\alpha_1}^{\beta_2}\,T(\alpha_1
p_1)\right]\delta(p_2+q_1)\,\delta(p_1-q_2)\nonu &+&S(\alpha_1
p_1-\beta_1 q_1)\left[
\delta_{\alpha_2}^{\beta_1}+\epsilon_{\alpha_2}^{\beta_1}\,T(\alpha_2
p_2)\right]\left[\delta_{\alpha_1}^{\beta_2}\,R(\alpha_1
p_1)\right]\delta(p_2-q_1)\,\delta(p_1+q_2)\nonu &+&S(\alpha_1
p_1-\beta_1 q_1)\left[\delta_{\alpha_2}^{\beta_1}\,R(\alpha_2
p_2)\right]\left[\delta_{\alpha_1}^{\beta_2}\,R(\alpha_1
p_1)\right]\delta(p_2+q_1)\,\delta(p_1+q_2)\nonu &+&\left[
\delta_{\alpha_1}^{\beta_1}+S(\alpha_1 p_1-\beta_2 q_2)S(\alpha_1
p_1+\beta_2 q_2)\epsilon_{\alpha_1}^{\beta_1}\,T(\alpha_1
p_1)\right]\left[
\delta_{\alpha_2}^{\beta_2}+\epsilon_{\alpha_2}^{\beta_2}\,T(\alpha_2
p_2)\right]\delta(p_1-q_1)\,\delta(p_2-q_2)\nonu &+&S(\alpha_1
p_1-\beta_2 q_2)\,S(\alpha_1 p_1+\beta_2
q_2)\left[\delta_{\alpha_1}^{\beta_1}\,R(\alpha_1
p_1)\right]\left[
\delta_{\alpha_2}^{\beta_2}+\epsilon_{\alpha_2}^{\beta_2}\,T(\alpha_2
p_2)\right]\delta(p_1+q_1)\,\delta(p_2-q_2)\nonu &+&\left[
\delta_{\alpha_1}^{\beta_1}+S(\alpha_1 p_1-\beta_2
q_2)\,S(\alpha_1 p_1+\beta_2
q_2)\epsilon_{\alpha_1}^{\beta_1}\,T(\alpha_1
p_1)\right]\left[\delta_{\alpha_2}^{\beta_2}\,R(\alpha_2
p_2)\right]\delta(p_1-q_1)\,\delta(p_2+q_2)\nonu &+&S(\alpha_1
p_1-\beta_2 q_2)\,S(\alpha_1 p_1+\beta_2
q_2)\left[\delta_{\alpha_1}^{\beta_1}\,R(\alpha_1
p_1)\right]\left[\delta_{\alpha_2}^{\beta_2}\,R(\alpha_2
p_2)\right]\delta(p_1+q_1)\,\delta(p_2+q_2)
\end{eqnarray}
We note that the result for the two-particle correlator differs
from that obtained in \cite{Mintchev:2002zd} by the appearance of two $S$
coefficients in the four last terms. This is due to the fact that
we started with a more general RT algebra where the defect
generators do not necessarily obey the linear relations used in
\cite{Mintchev:2002zd}. For the one-particle amplitudes, there are two
possibilities according to the relative signs of the $in$ and
$out$ states
\begin{equation}
^{out}\langle \mh_\pm,\mg_\pm\rangle^{in}=
\begin{cases}
\displaystyle\int_0^\infty \frac{dp}{2\pi}\,
\overline{\mh}_+(p)R(p)\mg_+(-p)\, ,\\
{}\\
\displaystyle\int_{-\infty}^0
\frac{dp}{2\pi}\,\overline{\mh}_-(p)R(-p)\mg_-(-p)\, ,
\end{cases}
\end{equation}
\begin{equation}
^{out}\langle \mh_\pm|\mg_\mp\rangle^{in}=
\begin{cases}
\displaystyle\int_0^\infty \frac{dp}{2\pi}\,
\overline{\mh}_+(p)T(p)\mg_-(p)\, ,\\
{}\\
  \displaystyle\int_{-\infty}^0
\frac{dp}{2\pi}\,\overline{\mh}_-(p)T(-p)\mg_+(p)\, .
\end{cases}
\end{equation}
One clearly sees the particle-impurity interaction through the
reflection coefficient $R$ for a final and an initial state on the
same half-line and through the transmission coefficient $T$
otherwise, as expected.
The particle-particle interaction through the bulk interaction
coefficient $S$ shows up in the $2^4$ different two-particle
amplitudes. As an illustration, we compute four such amplitudes
gathered into two generic expressions:
$$^{out}\langle
\mh_{1,\pm};\mh_{2,\pm}|\mg_{1,\pm};\mg_{2,\pm}\rangle^{in}=$$
\begin{eqnarray}
\int_{\RR^\pm}\frac{dp_1}{2\pi}\int_{\RR^\pm}\frac{dp_2}{2\pi}
\Big( \overline{\mh}_{1,\pm}(p_1)\overline{\mh}_{2,\pm}(p_2)R(\pm
p_2)S(\pm p_1 \pm p_2)R(\pm
p_1)\mg_{1,\pm}(-p_2)\mg_{2,\pm}(-p_1)\nonu
+\overline{\mh}_{1,\pm}(p_1)\overline{\mh}_{2,\pm}(p_2)R(\pm
p_1)S(\pm p_1 \pm p_2)S(\pm p_1 \mp p_2)R(\pm
p_2)\mg_{1,\pm}(-p_1)\mg_{2,\pm}(-p_2)\Big)
\end{eqnarray}
and
$$^{out}\langle
\mh_{1,\pm};\mh_{2,\pm}|\mg_{1,\pm};\mg_{2,\mp}\rangle^{in}=$$
\begin{eqnarray}
\int_{\RR^\pm}\frac{dp_1}{2\pi}\int_{\RR^\pm}\frac{dp_2}{2\pi}
\Big( \overline{\mh}_{1,\pm}(p_1)\overline{\mh}_{2,\pm}(p_2)R(\pm
p_2)S(\pm p_1 \pm p_2)T(\pm
p_1)\mg_{1,\pm}(-p_2)\mg_{2,\mp}(p_1)\nonu
+\overline{\mh}_{1,\pm}(p_1)\overline{\mh}_{2,\pm}(p_2)R(\pm
p_1)S(\pm p_1 \pm p_2)S(\pm p_1 \mp p_2)T(\pm
p_2)\mg_{1,\pm}(-p_1)\mg_{2,\mp}(p_2)\Big)
\end{eqnarray}
  More complex
transition amplitudes contain the same building blocks namely $R$,
$T$ and $S$, which shows that the corresponding processes involve
a succession of particle-impurity and particle-particle
interactions as expected from the factorized scattering occurring
in this integrable model.

\section{Discussion and conclusions \label{sect5}}

We have analyzed above the NLS model interacting with a $\delta$--type
impurity, establising the exact classical and quantum solutions. We
have shown that
an appropriate RT algebra and its Fock representation
allow to construct not only the scattering operator, but also
the off--shell quantum field $\Phi(t,x)$. As already mentioned in the
introduction, these results can be extended \cite{NLSlet}
to a whole class of point--like defects, substituting
(\ref{continuity},\ref{jump}) by the impurity
boundary  conditions
\begin{equation}
\lim_{x \downarrow 0}
\left(\begin{array}{cc} \langle \varphi \, ,\, \Phi (t,x) \psi \rangle  \\
\prt_x \langle \varphi \, ,\, \Phi (t,x) \psi \rangle
\end{array}\right) = \alpha
\left(\begin{array}{cc} a & b\\ c&d\end{array}\right)
\lim_{x \uparrow 0}
\left(\begin{array}{cc} \langle \varphi \, ,\, \Phi (t,x) \psi \rangle  \\
\prt_x \langle \varphi \, ,\, \Phi (t,x) \psi \rangle \end{array}\right) \, ,
\label{qbc}
\end{equation}
where
\begin{equation}
\{a,...,d \in \RR,\, \alpha \in \CC\, :\, ad -bc = 1,\,
{\overline \alpha} \alpha = 1, \} \, .
\label{parameters}
\end{equation}
In absence of impurity bound states, namely in the domain
\begin{equation}
\left\{\begin{array}{cc}
a+d+\sqrt{(a-d)^2+4} \leq 0 \, ,
& \quad \mbox{$b<0$}\, ,\\[1ex]
c (a+d)^{-1} \geq 0\, ,
& \quad \mbox{$b=0$}\, ,\\[1ex]
a+d-\sqrt{(a-d)^2+4} \geq 0 \, ,
& \quad \mbox{$b>0$}\, , \\[1ex]
\end{array} \right.
\label{nobs}
\end{equation}
one can treat the model closely following the $\delta$--impurity case,
because the corresponding reflection and transmission matrices
$\cal R$ and $\cal T$ have the same analytic properties as (\ref{forme-TR}).

We would like to comment finally on the symmetry content of
the solution derived in the paper.
It is quite obvious that impurities break down Galilean
(Lorentz) invariance of the {\it total} scattering matrix ${\bf S}$.
However, since the {\it bulk} scattering matrix $\cs$
describes the scattering away from the
impurity, some authors \cite{Delfino:1994nr}--\cite{Castro-Alvaredo:2002fc}
have assumed that $\cs$ preserves these symmetries and that the breaking in
${\bf S}$ is generated exclusively by the reflection and
transmission coefficients $\cal R$ and $\cal T$.
This assumption however, combined with the
conditions of factorized scattering, implies
\cite{Delfino:1994nr, Castro-Alvaredo:2002fc} that
$\cs$ is constant, which is too restrictive. In fact, one is left
with a few systems of limited physical interest.
In order to avoid this negative result, a consistent factorized
scattering theory
was developed in \cite{Mintchev:2002zd, Mintchev:2003ue},
which does not necessarily assume that $\cs$ is Galilean (Lorentz) invariant.
Since the impurity NLS model considered above is the first concrete application
of this framework with non--trivial bulk scattering, the lesson from
it is quite
instructive. Focusing on $\cs$ (\ref{bulks}), we see that Galilean
invariance is broken by the entries which
describe the scattering of two incoming particles
localized for $t\to - \infty$ on the different half-lines
$\RR_-$ and $\RR_+$ respectively. Indeed,
these entries depend on $k_1+k_2$ and not on $k_1-k_2$. An intuitive
explanation for this breaking is that before such particles scatter,
one of them
must necessarily cross the impurity. The non--trivial transmission is
therefore the origin of the symmetry breaking in $\cs$.
This conclusion agrees with the observation that in systems
which allow only reflection (e.g. models on the half-line), one can have
\cite{Cherednik:jt}--\cite{Liguori:1998xr} both Galilean (Lorentz) invariant
and non-constant bulk scattering matrices.

The issue of internal symmetries in the presence of impurities
has been
partially addressed in \cite{Mintchev:2003kh, varna03}.
In particular, the role
of the reflection and transmission elements of the RT algebra as
symmetry generators has been established. However, this
question deserves further investigation.
It will be interesting in this respect
to extend the analysis \cite{Mintchev:2001aq} of the
$SU(N)$--NLS model on
the half-line to the impurity case. Work is in progress on this aspect.

Let us conclude by observing that the concept of RT algebra indeed represents
a powerful tool for solving the NLS model with impurities.
We are currently exploring the possibility to apply this algebraic framework
also to the quantization of other integrable systems with defects.

\section*{Appendix}
\appendix

\section{Proof of theorem \ref{boundary-conditions} \label{app-2}}

First, notice that (\ref{continu}) and (\ref{saut}) translate into
\begin{eqnarray}
\label{continu2}
\lim_{x\to 0^+}\{\Phi_+(t,x)-\Phi_-(t,-x)\}&=&0\, ,\\
\label{saut2} \lim_{x\to
0^+}\{(\partial_x\Phi_+)(t,x)-(\partial_x\Phi_-)(t,-x)
\}-2\eta\,\Phi(t,0)&=&0
\end{eqnarray}
which we are going to check order by order in the Rosales
expansion. The idea is to introduce the one-to-one correspondence
\begin{equation}
\label{one-one}
\beta_\pm(p)=\frac{1}{2}\{\lda_+(p)\pm\lda_-(-p)\},~p\in\RR
\end{equation}
and it is not difficult to check that
\begin{equation}
\label{prop-beta}
\beta_\alpha(p)=B_\alpha(p)\beta_\alpha(-p)~,~~
\text{with}~B_\alpha(p)
=\alpha\frac{p-i\alpha\eta}{p+i\eta}~,~~\alpha=\pm\, .
\end{equation}
Take $n=0$ corresponding to the linear problem. One gets
\begin{eqnarray*}
\displaystyle\lim_{x\to
0^+}\left\{\Phi^{(0)}_+(t,x)-\Phi^{(0)}_-(t,-x)\right\}
&=&\int_{\RR}\frac{dp}{2\pi}\beta_-(p)e^{-ip^2t}\, ,\\
\displaystyle\lim_{x\to
0^+}\left\{(\partial_x\Phi^{(0)}_+)(t,x)-(\partial_x\Phi^{(0)}_-)(t,-x)
\right\}-2\eta\,\Phi^{(0)}(t,0)
&=&\int_{\RR}\frac{dp}{2\pi}(ip-\eta)\beta_+(p)e^{-ip^2t}\,,
\end{eqnarray*}
which vanish using the properties (\ref{prop-beta}). It is
interesting to note that the time-dependent phase $e^{-ip^2t}$,
being even in $p$, does not play any role in the vanishing of the
previous expressions. It will be the same in the following as we
shall see.

{F}or $n\ge 1$, we start by changing variables in the Rosales
expansion according to
$(p_1,\ldots,p_n,q_n,\ldots,q_0)\to(k_1,\ldots,k_{2n-1},-k_{2n}
\ldots,-k_0)$ and we use the one-to-one correspondence
(\ref{one-one}) to rewrite the left-hand side of (\ref{continu2})
as
\begin{eqnarray}
&&\lim_{x\to0^+}\{\Phi^{(n)}_+(0,x)-\Phi^{(n)}_-(0,-x)\}\ =
\label{int} \sum_{\alpha_0,\ldots,\alpha_{2n}=\pm}
(1-\prod_{i=0}^{2n}\alpha_i)\times\nonu &&\times
\int_{\RR^{2n+1}}\prod_{i=0}^{2n}\frac{dk_i}{2\pi}~
\bbet_{\alpha_1}(k_1)\ldots\bbet_{\alpha_{2n-1}}(k_{2n-1})
\beta_{\alpha_{2n}}(-k_{2n})\ldots\beta_{\alpha_0}(-k_0)
\frac{e^{-i\sum_{j=0}^{2n} k_j^2
t}}{\prod_{j=1}^{2n}(k_j+k_{j-1})}\qquad\quad
\end{eqnarray}
In view of the linear case, we
"$B_\alpha$-symmetrize" the integrand of the previous integral for
each $k_i$. Introducing
\begin{equation}
B_\alpha^\sigma(p)=
\begin{cases}
1~,& \text{for} ~\sigma=+\, ,\\
B_\alpha(p)~,& \text{for}~ \sigma=-\, ,
\end{cases}
\end{equation}
this reads
\begin{eqnarray*}
\frac{1}{2^{2n+1}}\sum_{\sigma_0,\ldots,\sigma_{2n}=\pm}
\frac{B^{\sigma_1}_{\alpha_1}(k_1)\ldots
B^{\sigma_{2n-1}}_{\alpha_{2n-1}}(k_{2n-1})
B^{\sigma_{2n}}_{\alpha_{2n}}(-k_{2n})\ldots
B^{\sigma_{0}}_{\alpha_{0}}(-k_{0})}
{\prod_{j=1}^{2n}(\sigma_j k_j+\sigma_{j-1} k_{j-1})}\\
\times \bbet_{\alpha_1}(k_1)\ldots\bbet_{\alpha_{2n-1}}(k_{2n-1})
\beta_{\alpha_{2n}}(-k_{2n})\ldots\beta_{\alpha_0}(-k_0)~
e^{-i\sum_{j=0}^{2n} k_j^2 t}
\end{eqnarray*}
which we rewrite as
\begin{eqnarray*}
\frac{1}{2^{2n+1}}\sum_{\sigma_0,\ldots,\sigma_{2n}=\pm}
B^{\sigma_1}_{\alpha_1}(k_1)\ldots
B^{\sigma_{2n-1}}_{\alpha_{2n-1}}(k_{2n-1})
B^{\sigma_{2n}}_{\alpha_{2n}}(-k_{2n})\ldots
B^{\sigma_{0}}_{\alpha_{0}}(-k_{0}) \prod_{j=1}^{2n}(\sigma_{j-1}
k_{j-1}-\sigma_j k_j)\\
\times
\frac{\bbet_{\alpha_1}(k_1)\ldots\bbet_{\alpha_{2n-1}}(k_{2n-1})
\beta_{\alpha_{2n}}(-k_{2n})\ldots\beta_{\alpha_0}(-k_0)}
{\prod_{j=1}^{2n}(k_{j-1}^2-k^2_j)} ~e^{-i\sum_{j=0}^{2n} k_j^2 t}
\end{eqnarray*}
Let us concentrate on the part depending on the $\sigma$'s.
Developing explicitly the
sum over $\sigma_{2n}$, one gets
\begin{eqnarray*}
\frac{1}{2^{2n+1}}\sum_{\sigma_0,\ldots,\sigma_{2n-1}=\pm}
B^{\sigma_1}_{\alpha_1}(k_1)\ldots
B^{\sigma_{2n-1}}_{\alpha_{2n-1}}(k_{2n-1})
B^{\sigma_{2n-2}}_{\alpha_{2n-2}}(-k_{2n-2})\ldots
B^{\sigma_{0}}_{\alpha_{0}}(-k_{0})
\\
\times\prod_{j=1}^{2n-1}(\sigma_{j-1} k_{j-1}-\sigma_j k_j)
\left(\delta_{\alpha_{2n},+}~\frac{2k_{2n}}{k_{2n}+i\eta}-
\delta_{\alpha_{2n},-}~2k_{2n}
\right)
\end{eqnarray*}
Collecting all the pieces depending on $k_{2n}$, one gets a
function proportional to
\begin{equation}
\label{special_fct}
\frac{k_{2n}}{k^2_{2n-2}-k^2_{2n}}
\left(\frac{\beta_+(-k_{2n})}{k_{2n}+i\eta}~
-\beta_-(-k_{2n})\right)\, .
\end{equation}
Now taking $\mu_+,\mu_-$ as in (\ref{form_mu}) it is not hard to
see that the function in brackets in (\ref{special_fct}) is
identically zero, implying the vanishing of (\ref{int}).

The case of the jump condition is treated in complete analogy.
Indeed, in evaluating the term proportional to $\eta$ in
(\ref{saut2}) in terms of $\beta_\pm$, all one has to do is to
replace $(1-\prod_{i=0}^{2n}\alpha_i)$ in (\ref{int}) by
$(1+\prod_{i=0}^{2n}\alpha_i)$. The rest of the argument implies
therefore that
\begin{equation}
\label{phi-n-zero} \Phi^{(n)}(0,0)=0~,~~n\ge 1\, .
\end{equation}
As for the term involving derivatives of the field, an analogous
treatment produces the following integrand
\begin{eqnarray*}
\frac{1}{2^{2n+1}}\sum_{\sigma_0,\ldots,\sigma_{2n}=\pm}
B^{\sigma_1}_{\alpha_1}(k_1)\ldots
B^{\sigma_{2n-1}}_{\alpha_{2n-1}}(k_{2n-1})
B^{\sigma_{2n}}_{\alpha_{2n}}(-k_{2n})\ldots
B^{\sigma_{0}}_{\alpha_{0}}(-k_{0}) \prod_{j=1}^{2n}(\sigma_{j-1}
k_{j-1}-\sigma_j k_j)
\\
\times\left(\sum_{j=0}^{2n}i\sigma_j k_j\right)
\frac{\bbet_{\alpha_1}(k_1)\ldots\bbet_{\alpha_{2n-1}}(k_{2n-1})
\beta_{\alpha_{2n}}(-k_{2n})\ldots\beta_{\alpha_0}(-k_0)}
{\prod_{j=1}^{2n}(k_{j-1}^2-k^2_j)} ~e^{-i\sum_{j=0}^{2n} k_j^2 t}
\end{eqnarray*}
This time, one has to develop the sum for $\sigma_{2n}$ and
$\sigma_{2n-1}$. This produces the function (\ref{special_fct})
but in the variable $k_{2n-1}$ and we know it vanishes. This leads
to
\begin{equation}
\label{deriv-n-zero} \lim_{x\to
0^+}\{(\partial_x\Phi^{(n)}_+)(0,x)-(\partial_x\Phi^{(n)}_-)(0,-x)
\}=0~,~~n\ge 1\, .
\end{equation}

As already mentioned, we see that the continuity and the jump
condition of the field hold for any time $t$. Put another way,
they are conserved in time and this is due to the dispersion
relation of the free Schr\"odinger equation (being quadratic in
$k_j$, it is not affected by all the symmetrizations $k_j\to -k_j$
involved in the proof).

It is remarkable that the jump condition actually decouples for
the nonlinear terms ($n\ge 1$) as seen from (\ref{phi-n-zero}) and
(\ref{deriv-n-zero}). This is also true for the continuity which,
combined with (\ref{phi-n-zero}) shows that
\begin{equation*}
\Phi_-^{(n)}(0,0)=\Phi_+^{(n)}(0,0)=0~,~~n\ge 1\, .
\qquad\qquad\qquad\qquad\rule{5pt}{5pt}
\end{equation*}

\section{Explicit form of the action of the creation
operator\label{app-3}}
The projector $P^{(n)}$ is constructed in \cite{Mintchev:2003ue} in terms of
the generators of the Weyl group associated to the root system of
the classical Lie algebra $B_n$ and of their representation on
$\cl^{\otimes n}$. In our context, we get for $\mf\in\cc$ and
$\varphi^{(n-1)}\in\ch^{(n-1)}$
\begin{eqnarray}
\left [\adag(\mf) \varphi \right ]_{\alpha_1\cdots
\alpha_{n}}^{(n)}(p_1,...,p_{n}) =\frac{1}{2\sqrt{n}}\sum_{k=1}^n
S(\alpha_{k-1}p_{k-1}-\alpha_k p_k)\ldots S(\alpha_1 p_1-\alpha_k
p_k)\nonu \times\Big( f_{\alpha_k}(p_k)+C_k(\alpha_1
p_1,\ldots,\alpha_n p_n)\left[T(\alpha_k
p_k)\mf_{-\alpha_k}(p_k)+R(\alpha_k
p_k)\mf_{\alpha_k}(-p_k)\right]\Big)\nonu\times
\varphi^{(n-1)}_{\alpha_1\ldots\hat{\alpha_k}\ldots\alpha_n}
(p_1,\ldots,p_n)
\end{eqnarray}
where we have defined
\begin{eqnarray*}
C_k(p_1,\ldots,p_n)=S( p_k-p_1)\ldots \widehat{S(p_k- p_k)}
\ldots S(p_k-p_n)\nonu \times
S(p_n+p_k)\ldots \widehat{S(p_k+p_k)}\ldots S(p_1+p_k)
\end{eqnarray*}
All the hatted symbols must be omitted.\\
One recognizes the reflected and transmitted structure inside the
square brackets which, combined with all the $S$ matrices, ensures
the properties (\ref{RT-prop})-(\ref{S-prop}) required for the
functions of $\ch^{(n)}$.


\end{document}